\documentclass[a4paper,fleqn]{cas-sc}

\usepackage[numbers]{natbib}

\input{1_tex/00_myConfig}

\def\tsc#1{\csdef{#1}{\textsc{\lowercase{#1}}\xspace}}
\tsc{WGM}
\tsc{QE}
\begin{document}
\let\WriteBookmarks\relax
\def\floatpagepagefraction{1}
\def\textpagefraction{.001}

% Short title
\shorttitle{Variability in HMI Experience}    

% Short author
\shortauthors{S. Kille et~al.}  

% Main title of the paper
\title [mode = title]{The Role of Variability in Human–Machine Interaction Experience}  

% Title footnote mark
% eg: \tnotemark[1]
\tnotemark[] 

% Title footnote 1.
% eg: \tnotetext[1]{Title footnote text}
%\tnotetext[1]{} 

% First author
%
% Options: Use if required
% eg: \author[1,3]{Author Name}[type=editor,
%       style=chinese,
%       auid=000,
%       bioid=1,
%       prefix=Sir,
%       orcid=0000-0000-0000-0000,
%       facebook=<facebook id>,
%       twitter=<twitter id>,
%       linkedin=<linkedin id>,
%       gplus=<gplus id>]

\author[1]{Sean Kille}[orcid=0000-0001-7550-7250]

% Corresponding author indication
\cormark[1]

% Footnote of the first author
%\fnmark[]

% Email id of the first author
\ead{sean.kille@kit.edu}

% URL of the first author
%\ead[url]{}

% Credit authorship
% eg: \credit{Conceptualization of this study, Methodology, Software}
\credit{}

% Address/affiliation
\affiliation[1]{organization={Institute of Control systems (IRS), Karlsruhe Institute of Technology (KIT)},
            addressline={Kaiserstr. 12}, 
            city={Karlsruhe},
            citysep={}, % Uncomment if no comma needed between city and postcode
            postcode={76131}, 
            state={BW},
            country={Germany}}

\author[1]{Jan Lars Hagemann}%[]

% Footnote of the second author
%\fnmark[]

% Email id of the second author
%\ead{}

% URL of the second author
%\ead[url]{}

% Credit authorship
\credit{}

% Address/affiliation
\affiliation[2]{organization={Cognition, Action, and Sustainability Unit, Department of Psychology, University of Freiburg},
            addressline={Engelbergerstr. 41}, 
            city={Freiburg},
            citysep={}, % Uncomment if no comma needed between city and postcode
            postcode={79085}, 
            state={BW},
            country={Germany}}

\author[2]{Anne Voormann}%[]

\author[1]{Balint Varga}%[]

\author[2]{Andrea Kiesel}%[]

\author[1]{Sören Hohmann}%[]

% Corresponding author text
\cortext[1]{Corresponding author}

% Footnote text
%\fntext[1]{}

% For a title note without a number/mark
%\nonumnote{}

% Here goes the abstract
\begin{abstract}
Human–machine interaction (HMI) requires control strategies that account for the nature of human motor behavior. Conventional shared-control and haptic-assistance methods typically ignore the stochastic nature of human behavior, potentially limiting both performance and human interaction experience. In this study, we designed an experimental setting and evaluated a novel human-variability-aware optimal controller. Participants performed a physically coupled haptic interaction task in three conditions: a controller mode that aims at conventionally reducing overall variability, a variability-aware controller mode designed to maintain human natural variability patterns, and a human-only control condition serving as a baseline. We analyzed behavioral variability, task performance, and human interaction experience. The results show that considering natural movement variability significantly increased perceived interaction quality in terms of usability while maintaining task performance. These findings highlight the importance of incorporating stochastic human movement characteristics into shared-control designs and demonstrate the feasibility and benefits of the proposed control strategy for human-centered control design of HMI. 
\end{abstract}

% Use if graphical abstract is present
%\begin{graphicalabstract}
%\includegraphics{}
%\end{graphicalabstract}

% Research highlights
% \begin{highlights}
% \item User study isolates effects of movement variability on interaction experience
% \item Preserving variability maintains performance while improving usability
% \item Motor variability structure shapes interaction experience in pHMI
% \end{highlights}

% Keywords
% Each keyword is seperated by \sep

\maketitle

% Main text

 % 1_tex/01_introduction.tex
\section{Introduction}

%\comment{Einreichung: Prio 1: Elsevier  Intl. Journal on Human-Computer Studies. Prio 2: Elsevier Computer in Human Behavior Reports? }

%\comment{todo: discuss title}

%TODOS Sean
%einheitliche Terminologie:
%--interaction mode / automation mode
%--positional variance / task-irrelevant variability

Physical human--machine interaction (pHMI) is increasingly shifting from loosely coupled assistance toward tightly integrated collaboration, in which humans and machines jointly generate motion and continuously exchange physical forces to achieve a shared task execution. Such tightly coupled systems aim to combine human adaptability and contextual flexibility with the precision, repeatability and endurance of machines~\cite{Lorenzini.2023, Farooq.2016a}---an essential capability in domains where full automation is infeasible, such as healthcare, rehabilitation, or skilled manual work~\cite{Kochhar.2023, Eloundou.2023}. 
Achieving this form of collaboration requires control strategies that not only improve task performance but also align with fundamental characteristics of human motor behavior and preserve a high-quality interaction experience. While objective task success is a necessary design criterion, it is not sufficient in tightly coupled human--machine systems, where users remain physically engaged and continuously regulate their actions in response to machine behavior. Systems that optimize performance at the expense of usability, perceived control, or experiential quality risk reduced trust, disengagement, and limited long-term acceptance. Particularly in domains such as healthcare, rehabilitation, or collaborative work, sustainable integration of assistive technologies depends not only on effectiveness, but on how interaction is experienced over repeated and prolonged use.

A central question, therefore, is which characteristics of human motor behavior must be respected in control design to sustain such high-quality interaction. In the current study, we focus on variability as a main characteristic of human motor behavior.

%Theoretische Einbettung: was ist human-centered design; welche anforderungen müssen dafür erfüllt werden (perceived agency, involvement, usability,..)
To achieve a high-quality interaction experience, human-centered design frameworks emphasize that interactive systems must support user involvement and sense of agency rather than optimize task performance alone~\cite{Bennett.2023,Schmager.2025}. This is because, in interactive settings, users remain active contributors whose ongoing actions and interpretations shape system behavior and outcomes. When users experience reduced involvement or diminished authorship over actions, they may disengage, miscalibrate their trust in the system, or ultimately reject it despite objectively successful task performance. Further research shows that perceived agency and autonomy significantly influence how users trust and engage with intelligent systems~\cite{FirminodeSouza.2025}. 

The sense of agency---the experience of being the author of one’s actions and their consequences---is commonly linked to the match between predicted and actual action outcomes~\cite{Haggard.2012}. According to comparator models of motor control, sense of agency is strengthened when sensory feedback aligns with internal predictions of movement consequences. Importantly, this predictive process is dynamic: post-hoc interpretations and contextual factors can shape how actions and outcomes are attributed~\cite{Johansson.2005}. Because internal predictions are formed based on repeated sensorimotor experience, they reflect the statistical structure of natural human movement, including its characteristic variability patterns.

From this perspective, control strategies that systematically alter or suppress natural movement variability may influence not only task performance but also how users predict and interpret their causal role within the interaction~\cite{Zanatto.2021}. In tightly coupled pHMI, where control policies are directly experienced through haptic forces, such effects may be particularly pronounced.
At the same time, this suggests that meaningful assistance does not necessarily undermine sense of agency, given that the supported action outcomes remain consistent with the user’s intentions and predictions.
Assistance that substantially reduces the need for active contribution may therefore diminish perceived authorship and experiential quality. 
Designing pHMI systems thus requires balancing task support with the preservation of the user’s active and self-attributed role in the collaborative process~\cite{Inga.2023}.

State-of-the-art shared-control and haptic-assistance frameworks typically rely on deterministic human models. Humans are commonly represented as mechanical impedance~\cite{Dong.2020} or as optimal controllers~\cite{Flad.2017, Varga.2024} acting to minimize a cost function under noise-free assumptions. While these model-based approaches have proven effective for anticipating human actions and improving task performance, they systematically neglect a key feature of human motor control: the stochastic nature of human motor behavior, which manifests as positional variability when performing repetitive movements. A broad body of motor control research shows that human movement is inherently stochastic, exhibiting structured variability characterized by low variance in task-relevant dimensions and systematically higher variance in task-irrelevant dimensions~\cite{Abend.1982, Harris.1998}. This structure reflects efficient allocation of control effort and sensorimotor precision and is well captured by linear–quadratic sensorimotor (LQS) models that incorporate both additive and signal-dependent noise~\cite{Todorov.2002,Todorov.2004}.

%todo Sean: hier potenziell sota-literatur, welche variabilität ermöglicht aber nicht systematisch untersucht, erwähnen. (Maeda.2017: ProMPs, Fitzsimons.2022: ergodic shared control framework)

Neglecting this structured stochasticity in control design may have implications that extend beyond motor performance. If the machine’s internal human model does not reflect the statistical structure of natural movement variability, a mismatch may arise between the user’s sensorimotor predictions and the forces imposed by the controller. Such mismatches can degrade task performance when assistance conflicts with the user’s natural coordination patterns. 
More importantly from a human-centered perspective, systematically constraining task-irrelevant variability may alter the sensorimotor contingencies through which users generate predictions and attribute action outcomes, thereby potentially degrading a person’s sense of agency and perceived usability. 
Such degradations are not merely subjective side effects: in tightly coupled pHMI, diminished agency or usability can influence trust calibration, engagement, and willingness to rely on the system, ultimately affecting how sustainably and effectively the human–machine partnership operates~\cite{Parasuraman.2000, Wen.2022}.
Related work in automation and human factors shows that both excessive and insufficient automation can reduce user involvement, perceived agency, and overall experience~\cite{Berberian.2019, Kaber.2004}. 

In tightly coupled haptic interaction, where humans continuously experience the machine’s behavior through physical forces, such effects may be amplified. Rather than seeking a single optimal level of automation that balances assistance and involvement, variability-respecting control offers the possibility to provide strong task support while preserving flexibility in task-irrelevant dimensions, thereby mitigating this trade-off by design. From the perspective of comparator models of agency, variability-respecting control can be interpreted as a design strategy that preserves the statistical regularities underlying users’ internal action predictions. Because users predictions are shaped by repeated experience with naturally structured variability, systematically suppressing task-irrelevant variability may alter the action outcomes to not match the users predictions, on which agency judgments rely. By contrast, preserving variability in task-irrelevant dimensions while supporting task-relevant dimensions maintains alignment between predicted and actual outcomes at the level of sensorimotor dynamics.  Variability-aware control can thus be theoretically motivated to not only be a refinement of human modeling, but as a principled approach to reconcile effective task support aiming to preserve experiential authorship in tightly coupled human--machine interaction, and thereby sustaining user involvement, perceived agency and usability.

\textit{Research gap.} Despite the well-established role of structured variability in motor control and the central importance of interaction experience in human-centered interaction design, the role of structured variability in pHMI research has not yet been explored. Existing shared-control frameworks predominantly evaluate performance improvements, while the experiential consequences of altering natural movement variability are not examined. In particular, it remains unclear whether systematically preserving task-irrelevant variability in tightly coupled haptic interaction influences human interaction experience, and whether such preservation can be achieved without sacrificing objective task performance. Consequently, the causal relationship between variability structure, task performance, and interaction experience in pHMI has not yet been empirically established.

\textit{Contribution.} To address this gap, we experimentally investigated how manipulating task-irrelevant movement variability influences both the task performance and the human experience of the interaction. In a controlled user study with a physically coupled haptic point-to-point task we compared three interaction modes:
\begin{enumerate}
    \item unsupported interaction (\textit{noSup})
    \item variability-constraining assistance (\textit{lowVar})
    \item variability-respecting assistance (\textit{highVar})
\end{enumerate}

% \begin{description}
%     \item[\textit{noSup}] A mode in which the human receives no support, in order to observe the natural behavior.
%     \item[\textit{lowVar}] A conventional controller mode that does not account to preserve natural variability.
%     \item[\textit{highVar}] A variability-aware controller mode designed to sustain natural human variability patterns.
% \end{description}
% \begin{itemize}
%     \item \textit{noSup}: a mode in which the human receives no support, in order to observe the natural behavior
%     \item \textit{lowVar}: a conventional controller mode that does not account to preserve natural variability, and
%     \item \textit{highVar}: a variability-aware controller mode designed to sustain natural human variability patterns.
% \end{itemize}
For the \textit{lowVar} and \textit{highVar} controller mode, we applied a recently proposed variability-aware optimal controller that aims to support the human in task-relevant dimensions while allowing a manipulation of variability in task-irrelevant dimensions to be either high or low~\cite{Kille.2024}. As a quantitative measure of variability, we used the maximum positional variance (maxVar) measured orthogonally to the task-relevant movement direction. 
In order to isolate the experiential effects of variability structure from differences in task success, the two assisted modes were parametrized to achieve comparable task-relevant performance while differing in their treatment of task-irrelevant variability. This design enables a controlled manipulation of movement variability under matched objective performance conditions. In addition to behavioral measures of task performance and variability, we assessed multiple facets of interaction experience, including sense of agency, self-efficacy, flow, and usability. By combining stochastic control design with a within-subject experimental protocol, the study allows us to empirically test whether variability structure itself influences subjective interaction quality. In doing so, we establish a causal link between task-irrelevant variability and human interaction experience in tightly coupled pHMI.

%offen: hier schon deutlich machen, dass wir max Var als task-irrelevant und endpointVar als task-relevant Maß nutzen? oder eher in materials and methods? 
% Anne: ich fände es schoen schon hier darüber zu lesen. Dann wuerde ich allerdings auch in dem Satz davor statt von naturally variable ones von task-irrelevant variability sprechen. Dann koenntet ihr den nachfolgenden Absatz auch ein wenig impliziter in die Theorie einfließen lassen. Der ließt sich fast schon als eine Beschreibung der Methoden, finde ich.

%We analyze behavioral variability, task performance, and human interaction experience to provide first empirical evidence on how respecting natural stochastic movement characteristics influences the human experience of collaboration.

 % 1_tex/02_method.tex

%%%%%%%%%%%%%%%%%%%%%%%%%%%%%%%%%%%%%%%%%%%%%%%%%%%%%%%%%%%%%%%%%%
%%%%%%%%%%%%%%%%%%%%%%%%%%%%%%%%%%%%%%%%%%%%%%%%%%%%%%%%%%%%%%%%%%
\section{Background}
%\comment{I would call this section "Background". I had a brief look into the latest publications of the journal and most use the term Background or theoretical review.}
%%%%%%%%%%%%%%%%%%%%%%%%%%%%%%%%%%%%%%%%%%%%%%%%%%%%%%%%%%%%%%%%%%
%%%%%%%%%%%%%%%%%%%%%%%%%%%%%%%%%%%%%%%%%%%%%%%%%%%%%%%%%%%%%%%%%%

%%%%%%%%%%%%%%%%%%%%%%%%%%%%%%%%%%%%%%%%%%%%%%%%%%%%%%%%%%%%%%%%%%
\subsection{Variability in human movement behavior}
%%%%%%%%%%%%%%%%%%%%%%%%%%%%%%%%%%%%%%%%%%%%%%%%%%%%%%%%%%%%%%%%%%
%\comment{Hier die menschliche natürliche Variabilität beschreiben. Verweis auf Abend82 (Trajectory variability), Engel01 (velocity-dependend variability) und Todorov02/04 (LQS Modellbeschreibung). Manifold theory (Scholz).  Ohne Formeln; evt systemmodell}

Human movement behavior is characterized by inherent variability, which can be observed even when individuals attempt to perform identical tasks repeatedly. This variability manifests in multiple dimensions, including spatial trajectories, velocity profiles, and force exertions. Studies such as \cite{Abend.1982} have highlighted the presence of trajectory variability in point-to-point movements, demonstrating that even under controlled conditions, individuals exhibit differences in their movement paths. \cite{Engelbrecht.2001} further elucidated the relationship between movement velocity and variability, indicating that faster movements tend to exhibit greater variability. Additionally, \cite{Todorov.2002, Todorov.2004} introduced the LQS model, which provides a theoretical framework for understanding how the central nervous system optimally controls movements in the presence of noise and uncertainty, leading to variability in motor outputs. This inherent variability is not merely a byproduct of motor execution but is believed to play a functional role in motor learning and adaptation, allowing individuals to explore different movement strategies and optimize performance over time. 

From a motor control perspective, not all variability is equivalent. A central distinction concerns task-relevant and task-irrelevant (redundant) variability~\cite{Todorov.2004, Scholz.1999}. Task-relevant variability directly affects goal achievement, such as deviations that increase target error or destabilize interaction dynamics. In contrast, task-irrelevant variability reflects fluctuations within redundant degrees of freedom that do not compromise task success. According to optimal feedback control theory, the motor system selectively regulates variability, strongly correcting deviations that threaten task goals while tolerating variability in dimensions that do not influence task outcome~\cite{Todorov.2002}. 

In physically coupled human--machine interaction, this distinction becomes particularly important. Control strategies that uniformly suppress variability may inadvertently constrain task-irrelevant movement dimensions, thereby altering natural motor expression without measurable gains in task performance. A variability-aware shared controller should therefore differentiate between these variability components, preserving functional redundancy while stabilizing task-critical dimensions.

In the present work, we adopt an operational interpretation of movement variability grounded in movement geometry and task structure. For repeated point-to-point movements, the task goal is defined along the primary movement direction connecting start and target. Deviations orthogonal to this axis do not directly affect goal attainment during the movement phase and can therefore be interpreted as task-irrelevant (redundant) variability. We quantify this component as the across-repetition positional variance perpendicular to the intended movement direction, evaluated primarily in the mid-region between start and target, where lateral deviations do not compromise endpoint success.

In contrast, variability in the same orthogonal dimension becomes task-relevant near the target point, where lateral deviations influence endpoint stability and final positioning accuracy. We therefore distinguish between mid-movement orthogonal variance as task-irrelevant variability and orthogonal variance in the target region as task-relevant variability. This operational separation allows us to differentiate functional motor flexibility from variability that directly affects goal achievement.

%%%%%%%%%%%%%%%%%%%%%%%%%%%%%%%%%%%%%%%%%%%%%%%%%%%%%%%%%%%%%%%%%%
\subsection{Variability-aware shared control}
%%%%%%%%%%%%%%%%%%%%%%%%%%%%%%%%%%%%%%%%%%%%%%%%%%%%%%%%%%%%%%%%%%

Given that human motor variability reflects structured processes, an important question is how shared-control systems account for this variability in their control architecture.

Model-based control strategies in pHMI have traditionally represented the human partner as a deterministic subsystem. Common controllers model the human as an impedance element~\cite{Dong.2020} or as a rational, deterministic agent in differential game settings~\cite{Flad.2017}. Although these approaches provide powerful tools for stability and performance optimization, they typically neglect the stochastic nature of human motor behavior. As a consequence, interaction variability is frequently attenuated as an unintended side effect of control design.

Early attempts to incorporate uncertainty into shared control frameworks primarily treated variability as a disturbance to be mitigated. For instance, adaptive optimal assistance strategies have adjusted cost functions based on sensing or behavioral uncertainty to improve robustness~\cite{Medina.2015}. Likewise, impedance adaptation schemes have been employed to counteract fluctuations in human input in order to stabilize interaction dynamics~\cite{Gribovskaya.2011}. In these approaches, stochasticity is acknowledged, yet its role remains compensatory rather than constitutive.

More recent work has reconsidered variability as a functional component of motor behavior, particularly in rehabilitation and motor learning contexts. Task-based shared control frameworks have been proposed that recompute assistance online to permit natural deviations while maintaining task success~\cite{Fitzsimons.2020}. Experimental studies have further shown that promoting motor variability during robotic assistance can enhance learning outcomes in dynamic tasks~\cite{Ozen.2021}. In related lines of work, external perturbations have been deliberately introduced to stimulate adaptive responses and improve neuromuscular capacity~\cite{Rubino.2024}. These findings suggest that variability can serve as a resource rather than merely as noise.

Beyond training applications, distribution-based representations have emerged as a means to formally encode stochastic aspects of human motion. Probabilistic Movement Primitives (ProMPs) represent human trajectories as distributions and allow robots to reproduce and adapt to variability in collaborative tasks~\cite{Maeda.2017}. Similarly, ergodic shared control approaches align robot behavior with spatial information distributions derived from human motion, thereby accounting for stochastic movement structure in closed-loop interaction~\cite{Fitzsimons.2022}. Collectively, these approaches reflect a growing recognition that variability carries structured information that can be exploited for adaptive interaction.

Generally, reinforcement learning (RL) approaches have likewise been applied to shared-control and human--robot interaction settings, enabling controllers to adapt policies from interaction data and optimize task-related rewards~\cite{Padalkar.2025, Javdani.2015, Christen.2019}. During learning, RL methods inherently involve stochastic exploration and may yield variable interaction dynamics. However, stochasticity primarily serves algorithmic purposes---such as exploration or robustness---rather than explicitly modeling or preserving the structured variability of human motor behavior. Consequently, while RL-based controllers may generate variability in practice, they typically do not differentiate between task-relevant and task-irrelevant variability nor systematically investigate how shaping variability structure influences the human interaction experience.

Despite this progress, two limitations remain. First, most variability-aware control designs have been evaluated primarily with respect to task performance, stability, or motor learning outcomes. 
However, objective performance metrics do not capture how the interaction is experienced by the human partner. In tightly coupled pHMI, users remain actively engaged in the control loop, and experiential factors such as perceived agency and usability can influence trust, engagement, and long-term acceptance of the system.
The impact of preserving or shaping variability on the human experience of interaction has received no attention. Second, existing approaches typically preserve or leverage variability implicitly, without offering explicit control over task-irrelevant variability while maintaining comparable task performance metrics. Consequently, it remains unclear whether observed benefits arise from improved performance, altered variability structure, or both.

To address these limitations, Human Variability-Respect\-ing Optimal Control (HVROC)~\cite{Kille.2024} introduces an explicit stochastic model of human movement variability into the control design. By embedding state-dependent noise characteristics within the optimal controller, the resulting feedback gains systematically shape task-irrelevant variability without compromising task accuracy. This design enables controlled manipulation of variability levels under otherwise comparable task conditions, thereby providing a methodological foundation for investigating how variability influences human interaction experience.

%%%%%%%%%%%%%%%%%%%%%%%%%%%%%%%%%%%%%%%%%%%%%%%%%%%%%%%%%%%%%%%%%%
\subsection{Human interaction experience}
%%%%%%%%%%%%%%%%%%%%%%%%%%%%%%%%%%%%%%%%%%%%%%%%%%%%%%%%%%%%%%%%%%
%Sense of Agency, Self-efficacy, Flow, Usability

Human interaction experience is a multidimensional construct capturing how users perceive, interpret, and regulate their interaction with a human--machine system. In physically coupled interactions, and in particular under variability-aware control, experience measures are essential to assess whether a controller not only achieves a good task performance but also whether the system supports natural and comprehensible interactions. Variability-aware control strategies explicitly shape how system responses relate to a user’s moment-to-moment actions; therefore, their evaluation requires metrics that consider the human experience of interaction as comprehensively as possible. In order to cover multiple aspects of human interaction experience, we assess perceived sense of agency, self-efficacy, flow, and usability in addition to task performance.

Sense of agency captures the extent to which users experience themselves as the cause of actions and outcomes. This construct is particularly relevant in variability-aware systems, as preserving natural movement variability may alter the balance between the human's intentions and the system's interventions. Measuring sense of agency allows to assess whether movements under adaptive assistance remains transparent and self-attributed, rather than being perceived as externally imposed~\cite{Haggard.2012, Jenkins.2021, Wen.2022}.

Self-efficacy reflects users’ beliefs in their capability to successfully perform the task under a given control mode. In assisted interaction, control strategies that overly constrain behavior may reduce perceived competence, whereas variability-respecting assistance may reinforce users’ confidence in their own abilities. Measuring self-efficacy thus provides insight into how different control modes affect users’ perceived autonomy and long-term empowerment~\cite{Hinds.1998}.

Flow describes a state of focused engagement that emerges when the task demands match well with the individual’s skill. In the present context, flow serves as an indicator of how seamlessly users can interact with the system without excessive cognitive or motor regulation. Variability-aware control is hypothesized to facilitate flow by allowing users to move in ways that feel natural and continuously responsive, rather than requiring constant adaptation to rigid system dynamics~\cite{Csikszentmihalhi.1997}.

Usability provides a complementary, system-oriented perspective on the interaction experience. While sense of agency, self-efficacy, and flow capture subjective experiential states, usability assesses whether users perceive the system as understandable, learnable, and practically supportive for task execution. Thus, it ensures that potential experiential benefits of variability-aware control are not achieved at the cost of increased interaction complexity or reduced clarity~\cite{Laugwitz.2008}.

Taken together, these measures capture distinct, yet interrelated facets of the human perceived experience of the interaction, enabling a structured assessment of how variability-aware control influences perceived control, competence, engagement, and system acceptance.

%\Comment{von Balint: Wäre hier ein "Concluding Research Gap" nicht hilfreich? :) Sean: der ist eigentlich schon am Ende von Kapitel 1 (vor den Hypothesen) zu finden }

%\Comment{Balint: okay, dann passt es 95 \% was noch zu uberlegn ware: am Ende der 3 SoA Absatze kurze Absatz warum SoA nicht gut genug ist. super niederschwellig machen ;)}

%Sean: steht eigentlich schon drin: "Despite this progress, two limitations remain. First, most variability-aware control designs have been evaluated pri-marily with respect to task performance, stability, or motor learning outcomes. The impact of preserving or shaping variability on the human subjective experience of interaction has received comparatively little attention. Second, existing approaches typically preserve or leverage variability implic- itly, without offering explicit control over task-irrelevant variability while maintaining comparable task performance metrics. Consequently, it remains unclear whether observed benefits arise from improved performance, altered variabil-ity structure, or both"

 %%%%%%%%%%%%%%%%%%%%%%%%%%%%%%%%%%%%%%%%%%%%%%%%%%%%%%%%%%%%%%%%%%
%%%%%%%%%%%%%%%%%%%%%%%%%%%%%%%%%%%%%%%%%%%%%%%%%%%%%%%%%%%%%%%%%%
\section{Study design}
%%%%%%%%%%%%%%%%%%%%%%%%%%%%%%%%%%%%%%%%%%%%%%%%%%%%%%%%%%%%%%%%%%
%%%%%%%%%%%%%%%%%%%%%%%%%%%%%%%%%%%%%%%%%%%%%%%%%%%%%%%%%%%%%%%%%%

% \Comment{einheitliche Verwendung von Begriffen und Abkürzungen:}
% \begin{itemize}
%     \item  variability-respecting control (variability-aware control) ...
%     \item key movements: die vier gecroppten bewegeungen, welche wir analysieren (hnl, hnr,...)
%     \item task-irrelevant variability: wird als allgemeiner Term in Introduction und Summary verwendet. Das Maß in diesem Paper dafür ist maximum (positional) variance (wird - wenn möglich - ausgeschrieben und nur in einzelnen plots (e.g. Fig.4) zu maxVar abgekürzt) 
%     \item task-relevant variability: wird als allgemeiner Term in Introduction und Summary verwendet. Das Maß in diesem Paper dafür ist  endpoint (positional) variance (wird - wenn möglich - ausgeschrieben und nur in vereinzelten Plots (e.g. Fig.4) zu endVar abgekürzt) 
%     \item weitere performanz-maße: endpoint error (keine abkürzung)
%     \item task measures vs experience measures (nicht: objective und subjective)
%     \item SoA, QUEAD sowie PU, PEU und E nicht abkürzen. (Sean: nicht von "emotions" sprechen, sondern von "emotional response". 
% \end{itemize}

The study within this paper investigates how preserving task-irrelevant variability during human--machine collaboration affects task performance and human interaction experience. We implemented a shared-control setup in which a robotic manipulator is virtually coupled to a simulated system. Both the human and an optional automation---the functional realization of the machine within this study---provide control inputs to the simulation.

%%%%%%%%%%%%%%%%%%%%%%%%%%%%%%%%%%%%%%%%%%%%%%%%%%%%%%%%%%%%%%%%%%
\subsection{Hypotheses}
%%%%%%%%%%%%%%%%%%%%%%%%%%%%%%%%%%%%%%%%%%%%%%%%%%%%%%%%%%%%%%%%%%

This study investigates how variability-aware control influences interaction experience and collaborative performance in an isolated point-to-point task. To this end, three automation modes were implemented which serve as the independent variable:

\begin{description}
    \item[\textit{noSup}] A mode without automation support, used to observe natural human behavior.
    \item[\textit{lowVar}] A mode that constrains natural human variability, representative of conventional controller designs.
    \item[\textit{highVar}] A variability-aware controller mode designed to sustain natural human movement variability.
\end{description}

% \begin{itemize}
%     \item \textit{noSup}: a mode without automation support, used to observe natural human behavior,
%     \item \textit{lowVar}: a mode that constrains natural human variability, representative of conventional controller designs,
%     \item \textit{highVar}: a variability-aware controller mode designed to sustain natural human movement variability.
% \end{itemize}

The \textit{noSup} mode serves as a baseline condition for capturing each participant’s natural movement patterns, movement variability, and interaction experience. The effects of the \textit{lowVar} and \textit{highVar} modes are compared relative to this baseline.

To structure the analysis, the study is guided by the following hypotheses:
\begin{description}
    \item[H1:] The \textit{highVar} mode yields higher task-irrelevant variability than the \textit{lowVar} mode.
    
    \item[H2a:] Task-relevant performance in \textit{highVar} exceeds that of \textit{noSup}.
    
    \item[H2b:] Task-relevant performance does not differ between \textit{highVar} and \textit{lowVar}.
    
    \item[H3:] The \textit{highVar} mode results in higher ratings of interaction experience than both \textit{noSup} and \textit{lowVar}.
\end{description}

% \begin{itemize}
%     \item H1: The \textit{highVar} mode leads to higher task-irrelevant variability than the \textit{lowVar} mode.
%     \item H2a: Task-relevant performance in \textit{highVar} is higher than in \textit{noSup} mode. 
%     \item H2b: Task-relevant performance does not differ between the \textit{highVar} and \textit{lowVar} modes. 
%     \item H3: The \textit{highVar} mode leads to higher ratings of interaction experience compared to both \textit{noSup} and \textit{lowVar}. %SoA, PU, PEU and flow. 
% \end{itemize}

The first hypothesis evaluates whether the control modes operate as intended by systematically manipulating task-irrelevant variability. To validate the effectiveness of \highvar mode, Hypothesis~H2a analyzes task-relevant performance compared to the baseline. Hypothesis~H2b is formulated as a non-inferiority hypothesis, reflecting the expectation that preserving task-irrelevant variability does not come at the expense of task-relevant performance.
The third hypothesis is based on the assumption, that preserving natural human movement variability leads to a more positive interaction experience.

%%%%%%%%%%%%%%%%%%%%%%%%%%%%%%%%%%%%%%%%%%%%%%%%%%%%%%%%%%%%%%%%%%
\subsection{Shared-control haptic interface}
%%%%%%%%%%%%%%%%%%%%%%%%%%%%%%%%%%%%%%%%%%%%%%%%%%%%%%%%%%%%%%%%%%

%%%%%%%%%%%%%%%%%%%%%%%%%%%%%%%%%%%%%%%%%%%%%%%%%%%%%%%%%%%%%%%%%%
\subsubsection{KUKA haptic interface}
\begin{figure}
    \begin{center}
    \includegraphics[width=6.4cm]{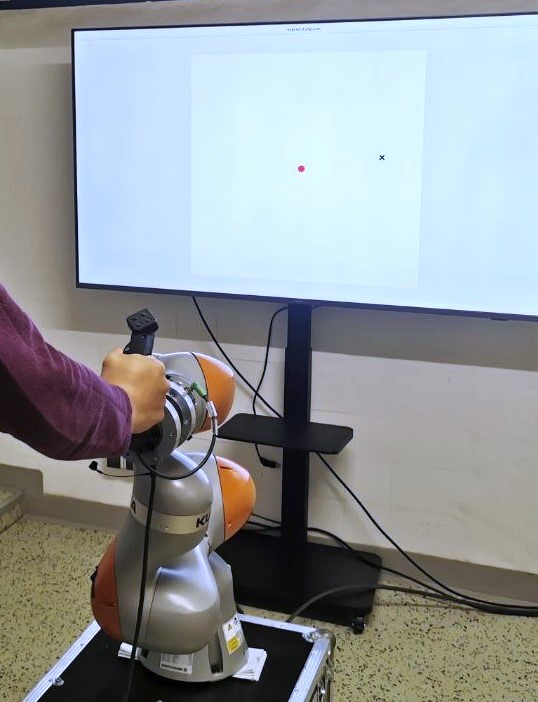}    % The printed column width is 8.4 cm.
    \caption{Experimental setup.} 
    \label{fig:setup}
    \end{center}
\end{figure}

We developed an experimental setup for shared-control haptic interaction in which a human user and an optional automation jointly control a simulated system in a 2D workspace (motion constrained to a fixed height). The haptic interface is a KUKA LBR iiwa 14 R820 robotic arm (KUKA Deutschland GmbH, Augsburg, Germany), following the setup in~\cite{Braun.2023}. A force-torque sensor (Schunk SE \& Co. KG, Lauffen/Neckar, Germany) mounted at the end-effector records interaction forces (and torques) at \SI{70}{\hertz}. A graphical user interface (GUI) displays the current end-effector and target position. Human and automation inputs are applied as forces to the simulation. The simulation is coupled to the end-effector via an impedance controller that renders haptic feedback to the user. All software components (GUI, automation, and target generation) run in ROS2~\cite{Macenski.2022} at \SI{5}{\milli\second} cycle time. 
The complete experimental setup is illustrated in Figure~\ref{fig:setup}.

%tbd: Jan: Systemaufbau von Abbildung Shared-Control?

%%%%%%%%%%%%%%%%%%%%%%%%%%%%%%%%%%%%%%%%%%%%%%%%%%%%%%%%%%%%%%%%%%
\subsubsection{Simulated system dynamics}

The simulated system is modeled as a linear point-mass system with mass-damper dynamics. Its behavior is described by the following discrete-time linear state-space model:
\begin{align}
\boldsymbol{x}_{k+1}=
\begin{bmatrix}
\boldsymbol{A} & 0 \\
0 & \boldsymbol{A}
\end{bmatrix}
\boldsymbol{x}_{k}
+
\begin{bmatrix}
\boldsymbol{B} & 0 \\
0 & \boldsymbol{B}
\end{bmatrix}
(\boldsymbol{u}_{\mathrm{H}} + \boldsymbol{u}_{\mathrm{A}}).
\end{align}
The human and the automation apply additive input forces $\boldsymbol{u}_{\mathrm{A}}$ and $\boldsymbol{u}_{\mathrm{H}}$ to the system. The state vector is defined as $\bm{x}=\begin{bmatrix} p_x & \dot{p}_x & p_y & \dot{p}_y \end{bmatrix}^\top$, where $p_x$, $p_y$ denote position and $\dot{p}_x$, $\dot{p}_y$ velocity. The system matrix $\boldsymbol{A}$ and input matrix $\boldsymbol{B}$ describe the one-dimensional point-mass dynamics and are identical for both spatial dimensions:
\begin{equation}
\boldsymbol{A} = 
\begin{bmatrix}
1 & \Delta t \\
0 & 1 - \frac{d}{m} \Delta t
\end{bmatrix}, \quad
\boldsymbol{B} = 
\begin{bmatrix}
0 \\
\frac{\Delta t}{m}
\end{bmatrix}.
\end{equation}
The simulation runs at a time step of $\Delta t = \SI{5}{\milli\second}$, with mass $m = \SI{50}{\kilogram}$ and damping coefficient $d = \SI{75}{\kilo\gram\per\second}$.

%%%%%%%%%%%%%%%%%%%%%%%%%%%%%%%%%%%%%%%%%%%%%%%%%%%%%%%%%%%%%%%%%%
\subsubsection{Automation design}\label{sec:automation}
The automation computes the automation force $\boldsymbol{u}_{\mathrm{A},k}$ using a state-feedback control law of the form:
\begin{equation}
\boldsymbol{u}_{\mathrm{A},k} = -\boldsymbol{L}_k \boldsymbol{x}_{\mathrm{aug},k}.
\end{equation}
The control action is based on an augmented state vector that represents the position error $\boldsymbol{e}_{\mathrm{p}}$ and the Cartesian velocity $\boldsymbol{\dot{p}}$ separately for the $x$- and $y$-dimensions:
\begin{equation}
\boldsymbol{x}_{\mathrm{aug}} =
\begin{bmatrix}
e_{x} & \dot{p}_x & e_{y} & \dot{p}_y
\end{bmatrix}^{\top}.
\end{equation}

Two control modes were implemented. The \highvar mode is designed to preserve human movement variability, whereas the \lowvar mode minimizes variability. % via increased damping. 
To minimize confounding effects due to human reaction time, the automation was activated only when the human-applied force exceeded a predefined threshold of \SI{5}{\newton}.

The automation design is based on the previously introduced concept of Human-Variability-Respecting Optimal Control (HVROC)~\cite{Kille.2024}. HVROC formulates an optimization problem with cost terms that explicitly weight position variability. 
%The resulting optimal feedback law is time-dependent, exhibiting high gains at movement onset that decay toward zero over the course of the motion. 
    
In this work, an adjusted error-dependent approximation of the HVROC time-dependent feedback matrix is employed, to better handle the multitude of participants and their individual behavior. The feedback gains penalizing position error and velocity are expressed as a function of the Euclidean target distance
$e = \sqrt{e_{\mathrm{x}}^2 + e_{\mathrm{y}}^2}$.

The feedback matrix is defined as:
\begin{equation}
\boldsymbol{L}(e) = 
\begin{bmatrix}
L_{\mathrm{p,x}}(e) & L_{\mathrm{d,x}}(e) & L_{\mathrm{p,y}}(e) & L_{\mathrm{d,y}}(e)
\end{bmatrix},
\end{equation}
where $L_{\mathrm{p}}$ and $L_{\mathrm{d}}$ denote the position and velocity feedback gains, respectively. Identical gains are applied for both spatial dimensions.

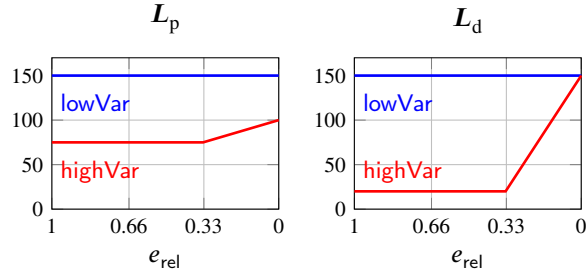
\begin{figure}
\centering
    \begin{tikzpicture}
  \begin{axis}[
   width=3cm,
   height=2cm,
   scale only axis,
   xlabel near ticks,
ylabel near ticks,
   at={(0cm,0cm)},
   xlabel={$e_{\text{rel}}$},
   x dir=reverse,
   xtick={0,0.33,0.66,1},
   xticklabel style={font=\footnotesize  },
   xlabel style={font=\normalsize },
   ytick={0,50,100,150},
   xmin=0, xmax=1,
   ymin=0, ymax=170,
yticklabel style={font=\footnotesize},
   ylabel style={font=\normalsize},
   title={$\boldsymbol{L}_{\mathrm{p}}$},
   title style={font=\normalsize \bfseries},
   grid=both,
   legend pos=south east,
]
   % einfache Linie mit Punkten
   \addplot[mark=None, blue, line width=1pt] coordinates {(1,150) (0,150)};
      \node[anchor=south west,blue, font=\small ] at (axis cs:1.0,100) {lowVar};
   \addplot[mark=None, red, line width=1pt] coordinates {(1,75) (1/3,75) (0,100)};
   \node[anchor=south west, red, font=\small ] at (axis cs:1.0,20) {highVar};

  \end{axis}
    \begin{axis}[
   width=3cm,
   height=2cm,
   scale only axis,
   xlabel near ticks,
    ylabel near ticks,
   at={(4cm,0cm)},
   xlabel={$e_{\text{rel}}$},
   x dir=reverse,
   xtick={0,0.33,0.66,1},
   xticklabel style={font=\footnotesize},
   xlabel style={font=\normalsize},
   ytick={0,50,100,150},
   xmin=0, xmax=1,
   ymin=0, ymax=170,
    yticklabel style={font=\footnotesize},
   ylabel style={font=\normalsize },
   title={$\boldsymbol{L}_{\mathrm{d}}$},
   title style={font=\normalsize \bfseries},
   grid=both,
   legend pos=south east,
  ]
    % einfache Linie mit Punkten
      \addplot[mark=None, blue, line width=1pt] coordinates {(1,150) (0,150)};
      \node[anchor=south west,blue, font=\small ] at (axis cs:1.0,100) {lowVar};
     
      \addplot[mark=None, red, line width=1pt] coordinates {(1,20) (1/3,20) (0,150)};
      \node[anchor=south west, red, font=\small ] at (axis cs:1.0,20) {highVar};
  \end{axis}
\end{tikzpicture}   % The printed column width is 8.4 cm.
    \caption{Feedback matrices as a function of position error.} 
    \label{plot:automation_gains}
\end{figure}

In the \lowvar mode, a stationary HVROC solution with $L_{\mathrm{p}} = L_{\mathrm{d}} = 150$ is used. In contrast, the \highvar mode employs position-error-dependent gain scheduling, with feedback gains increasing as the system approaches the target. The gains are initially set to $L_{\mathrm{p}} = 75$ and $L_{\mathrm{d}} = 20$ and increase linearly and monotonically to $L_{\mathrm{p}} = 100$ and $L_{\mathrm{d}} = 150$ once the position error falls below:
\begin{equation}
e_{\mathrm{trigger}} =   1/3 \, e_{\mathrm{start}}.
\end{equation}
The position error at movement onset is denoted as $e_{\mathrm{start}}$. The state-feedback gains are shown in Fig.~\ref{plot:automation_gains} as a function of the relative position error $e_{\mathrm{rel}}$ normalized by $e_{\mathrm{start}}$.

This design yields low automation assistance over most of the movement to preserve human movement variability, while providing increased damping near the target point to ensure a stable and well-controlled deceleration in the \highvar mode.

%evt die aktuell in IV-A beschriebenen modes hier schon vorstellen und einführen, da die hypothesen in III-C bereits darauf verweisen

%%%%%%%%%%%%%%%%%%%%%%%%%%%%%%%%%%%%%%%%%%%%%%%%%%%%%%%%%%%%%%%%%%
\subsection{Experimental task}\label{sec:ExperimentalTask}
%%%%%%%%%%%%%%%%%%%%%%%%%%%%%%%%%%%%%%%%%%%%%%%%%%%%%%%%%%%%%%%%%%

The experimental task requires participants to perform point-to-point movements between four target points arranged in a rectangular layout. Participants received visual feedback via a GUI that displays the target points and the current end-effector position. Haptic feedback was provided via the KUKA haptic interface, allowing participants to perceive the forces generated by the shared control system in addition to the visual position feedback. Participants were instructed to perform the movements as fast and as accurate as possible, thereby balancing speed and precision.

Each participant completed 82 movements per mode in pseudorandomized order, including 12 repetitions of each of the four key movement directions (a–d) shown in Fig.~\ref{plot:targetpoints}.

%The task was deliberately designed to allow the expression of natural movement variability. By specifying only start and target positions while leaving the intermediate trajectory unconstrained, the task admits a large set of equally valid movement solutions. Participants were encouraged to balance speed and accuracy, which naturally gives rise to individual movement strategies and trial-to-trial variability. The absence of explicit path, timing, or kinematic constraints further prevents artificial regularization of the movement. This design ensures that observed variability reflects intentional and strategy-dependent motor behavior rather than task-induced constraints, making it suitable for studying how shared-control strategies influence the preservation or suppression of human movement variability.

The task design intentionally permits the expression of natural movement variability by specifying only start and target positions while leaving the movement trajectory unconstricted. This redundancy, combined with repeated trials and the instruction to emphasize speed and accuracy, allows individual movement strategies and variability to emerge without task-induced regularization.

%%%%%%%%%%%%%%%%%%%%%%%%%%%%%%%%%%%%%%%%%%%%%%%%%%%%%%%%%%%%%%%%%%
%%%%%%%%%%%%%%%%%%%%%%%%%%%%%%%%%%%%%%%%%%%%%%%%%%%%%%%%%%%%%%%%%%
\section{Study procedure}
%%%%%%%%%%%%%%%%%%%%%%%%%%%%%%%%%%%%%%%%%%%%%%%%%%%%%%%%%%%%%%%%%%
%%%%%%%%%%%%%%%%%%%%%%%%%%%%%%%%%%%%%%%%%%%%%%%%%%%%%%%%%%%%%%%%%%
\begin{figure}
    \begin{center}
    \definecolor{color_b}{rgb}{0.06600,0.44300,0.74500}%
\definecolor{color_c}{rgb}{0.86600,0.32900,0.00000}%
\definecolor{color_d}{rgb}{0.92900,0.69400,0.12500}%
\definecolor{color_a}{rgb}{0.52100,0.08600,0.81900}%
\definecolor{mycolor5}{rgb}{0.12941,0.12941,0.12941}%

\begin{tikzpicture}[x=1.7cm, y=1.7cm, font=\small]
   % Simples Koordinatensystem
      % Horizontal: p_y von -0.4 bis 0.4m
      % Vertikal: p_x von 0 bis -1m (negative Werte)
      %\draw[step=1cm,lightgray,very thin] (-2,0) grid (2,1.0);
      \draw[thick,->] (-2,-0.3) -- (2.1,-0.3) node[anchor=south] {\small $p_\mathrm{x}$ in m};
      \draw[thick,->] (-2,-0.3) -- (-2,1.2) node[anchor=north, yshift=12pt] {\footnotesize $p_{\mathrm{y}}$ in m};

      % Beschriftung p_y (horizontal) mit Dezimalwerten
      \foreach \x in {-0.15,-0.075,0,0.075,0.15}
         \draw (10*\x,-0.25) -- (10*\x,-0.35) node[anchor=north, yshift=0.1cm] {\footnotesize \x};

      % Beschriftung p_x (vertikal) mit negativen Dezimalwerten
      \foreach \y in {0.0,0.1,0.2}
         \draw (-1.95,5*\y) -- (-2.05,5*\y) node[anchor=east] {\footnotesize \y};
         
      % Punkte mit Beschriftung unterhalb
      \node[draw,circle,fill=black,inner sep=1.5pt,label=south:{\footnotesize $p_2$}] at (1.5,0) {};
      \node[draw,circle,fill=black,inner sep=1.5pt,label=south:{\footnotesize $p_1$}] at (-1.5,0) {};
      \node[draw,circle,fill=black,inner sep=1.5pt,label=north:{\footnotesize $p_3$}] at (1.5,1) {};
      \node[draw,circle,fill=black,inner sep=1.5pt,label=north:{\footnotesize $p_2$}] at (-1.5,1) {};
      %\node[draw,circle,fill=red,inner sep=2pt,label=below:{\small $p_0$}] at (0,0) {}; % Initial point

    % andere mögliche Bewegungen
    \draw[<->,shorten <=4pt,shorten >=4pt, gray] (-1.5,0.0) -- (1.5,1);
      \draw[<->,shorten <=4pt,shorten >=4pt, gray] (-1.5,1) -- (1.5,0);
      \draw[<->,shorten <=4pt,shorten >=4pt, gray] (-1.5,0) -- (-1.5,1);
      \draw[<->,shorten <=4pt,shorten >=4pt, gray] (1.5,0) -- (1.5,1);
      
      % key movements
      \draw[shorten <=4pt,shorten >=4pt, thick, arrows={->[harpoon]}, color_a] (-1.5,1.02) -- (1.5,1.02) node[midway,above,yshift=0.5pt] {\small \textit{a}};
    \draw[shorten <=4pt,shorten >=4pt, thick, arrows={->[harpoon]}, color_b] (1.5,0.98) -- (-1.5,0.98) node[midway,below,yshift=-0.5pt] {\small \textit{b}};

      \draw[shorten <=4pt,shorten >=4pt, thick, arrows={->[harpoon]}, color_c] (-1.5,0.02) -- (1.5,0.02) node[midway,above,yshift=-0.5pt] {\small \textit{c}};
      \draw[shorten <=4pt,shorten >=4pt, thick, arrows={->[harpoon]}, color_d] (1.5,-0.02) -- (-1.5,-0.02) node[midway,below,yshift=-0.5pt] {\small \textit{d}};
   \end{tikzpicture}   % The printed column width is 8.4 cm.
    \caption{Target points $p_{\mathrm{
    1, \dots,4}}$ with key movements a--d. } 
    \label{plot:targetpoints}
    \end{center}
%\end{figure}
%\begin{figure}
   \begin{center}
    \input{3_plots/one_subject.tex} 
    \caption{Derivation of position variance from one example subject: For each key movement a-d (top), the y-positional variance is calculated (middle) and averaged per timestep over the four key movements (bottom).  The maximum peak of this variance represents the task-irrelevant variability. Endpoint variability is calculated as mean variance over the gray area.}
    \label{plot:single_subject_trajectories}
    \end{center}
\end{figure}

%%%%%%%%%%%%%%%%%%%%%%%%%%%%%%%%%%%%%%%%%%%%%%%%%%%%%%%%%%%%%%%%%%
\subsection{Study protocol}\label{sec:studyProtocol}
%%%%%%%%%%%%%%%%%%%%%%%%%%%%%%%%%%%%%%%%%%%%%%%%%%%%%%%%%%%%%%%%%%

A total of 44 participants took part in the study, of which three were excluded due to technical problems or early study termination. Participants were recruited from two universities, specifically the Karlsruhe Institute of Technology as well as the University of Freiburg. 

At the beginning of the study, participants received a detailed introduction to the task, including an explanation of the experimental procedure and safety instructions. This was followed by a familiarization phase in which participants interacted with the system without automation support (\textit{noSup} mode). During this phase, participants performed a brief task sequence encompassing all movement types.

The three modes were presented in a pseudorandomized order to mitigate potential order effects and were announced to participants using neutral labels (Mode A–C), without revealing the underlying experimental condition. Each participant completed all three conditions.

After completing the task described in Section~\ref{sec:ExperimentalTask}, participants filled out a questionnaire capturing aspects of their interaction experience after each mode. Upon completion of all conditions, a final questionnaire was administered to collect demographic information. The entire experiment lasted approximately~\SI{50}{\min}. 
The experimental protocol was approved by the responsible Research Ethics Board. The protocol was carried out in accordance with the relevant guidelines and regulations. 
All participants provided written informed consent prior to participation.

% The participants performed interactions in three experimental conditions which serve as the independent variable:
% \begin{itemize}
%     \item \textbf{Human-only / No Support} (\textit{noSup}) mode:
%     \item \textbf{Low Variability Shared Control} (\textit{lowVar}) mode:
%     \item \textbf{High Variability Shared Control} (\textit{highVar}) mode:
% \end{itemize}

\subsection{Data processing}\label{sec:processing}

Prior to analysis, all recorded interaction data were preprocessed. 
To eliminate the influence of reaction time, the onset of each key movement trajectory was cropped based on a human-applied force threshold of \SI{5}{\newton}, corresponding to the point at which automation assistance was engaged (see Sec.~\ref{sec:automation}). This force-based onset definition ensured consistent alignment across participants and control modes. 

For each subject and condition, and for each key movement $k\in\mathcal{K}$ ($|\mathcal{K}|=K=4$), we considered $R=12$ within-subject repetitions.
All repetition trajectories were time-cropped to a common length $n=1000$, corresponding to a duration of \SI{5}{\second}. The chosen duration covered the full movement for all participants and conditions.
This ensures that index $i=1,\dots,n$ represents the same movement phase across repetitions.

%%%%%%

Let $\bm{p}_{i}^{(k,r)}$ denote the position at time step $i$ of repetition $r$ of key movement $k$.
For each participant and mode, and for each $k$ and time step $i$, the across-repetition mean was computed as
\begin{equation}
    \bar{p}_{\square,i}^{(k)} = \frac{1}{R}\sum_{r=1}^{R} p_{\square,i}^{(k,r)}, \qquad \square \in \{x,y\}.
\end{equation}
The mean position vector is then obtained by stacking the component-wise means:
\begin{equation}
    \bar{\bm{p}}_{i}^{(k)} =
    \begin{bmatrix}
    \bar{p}_{x,i}^{(k)}\\
    \bar{p}_{y,i}^{(k)}
    \end{bmatrix}.
\end{equation}

%%%%%%%%%%%%%%%%%%%%%%%%%%%%%%%%%%%%%%%%%%%%%%%%%%%%%%%%%%%%%%%%%%
\subsection{Measures for task performance and variability}%oder Performance Measures? allgemein: movement measures?
%%%%%%%%%%%%%%%%%%%%%%%%%%%%%%%%%%%%%%%%%%%%%%%%%%%%%%%%%%%%%%%%%%

To quantitatively evaluate the effect of the automation modes on task performance and human-machine behavior, the following task measures were computed, see Figure~\ref{plot:single_subject_trajectories}, based on the preprocessed data presented in \ref{sec:processing}.

\begin{description}
     \item[Task-irrelevant variability (Maximum Variance)] $\posvarmax$: \hfill \\  Based on the across-repetition mean $\bar{p}_{y,i}^{(k)}$, the time-resolved positional variance across the $R$ repetitions was computed for each key movement $k$ and time step $i$ as
\begin{equation}
    \mathrm{Var}_r\!\left(p_{y,i}^{(k)}\right)
    = \frac{1}{R-1}\sum_{r=1}^{R}\left(p_{y,i}^{(k,r)}-\bar{p}_{y,i}^{(k)}\right)^2 .
\end{equation}
The variance trajectories were then averaged across the $K=4$ key movements,
\begin{equation}
    \posvar(i) = \frac{1}{K}\sum_{k\in\mathcal{K}} \mathrm{Var}_r\!\left(p_{y,i}^{(k)}\right),
    \qquad i=1,\dots,n.
\end{equation}
    As a measure of task-irrelevant variability we define the \textbf{maximum} value of the averaged positional \textbf{variance}
    \begin{equation}
    \posvarmax = \max_{i\in\{1,\dots,n\}} \posvar(i)
\end{equation}
    over the movement duration. 
    The positional variance for each key movement is displayed in Fig.~\ref{plot:single_subject_trajectories} (middle), maximum value and the mean variance trajectory are displayed in Fig.~\ref{plot:single_subject_trajectories} (bottom).

    % \item[Task-irrelevant variability (Maximum Variance)] \hfill \\ $\posvarmax$: For each key movement k the positional variance for each time step $i=1,\dots,n$ was calculated:
    % \begin{equation}
    %     \mathrm{Var}\big(\boldsymbol{p}_{\mathrm{y}}\big)^{(k)}= \frac{1}{n-1}\sum^n_{i} \left( \boldsymbol{p}_{\mathrm{y},i}^{(k)} - \overline{\boldsymbol{p}}_{\mathrm{y}}^{(k)} \right)^2.
    % \end{equation}
    % Based on these, the averaged variance over the key movements was calculated:
    % \begin{equation}
    %     \posvar= \frac{1}{4} \sum_{k \in \{a,b,c,d\}} \mathrm{Var}\big(\boldsymbol{p}_{\mathrm{y}}^{(k)}\big).
    % \end{equation}
    % As a measure of task-irrelevant variability we define the \textbf{maximum} value of the averaged positional \textbf{variance} $\posvarmax$ over the four key movements. 
    % The positional variance for each key movement is displayed in Fig.~\ref{plot:single_subject_trajectories} (middle), maximum value and the mean variance trajectory are displayed in Fig.~\ref{plot:single_subject_trajectories} (bottom).
    
    \item[Settling time] $t_\mathrm{s}$:\hfill \\
The settling time is defined as the elapsed time until the movement enters and remains within a tolerance region of \SI{1.25}{\centi\meter} around the target point. For each key movement $k$, the settling time $t_\mathrm{s}$ is determined as mean over all repetitions and key movements:
% \begin{equation}
%     t_\mathrm{s} = \frac{1}{K} \sum_{k \in \mathcal{K}} t_\mathrm{s}^{(k)}.
% \end{equation}
\begin{equation}
    t_{\mathrm{s}}
    =
    \frac{1}{K R}
    \sum_{k \in \mathcal K}
    \sum_{r=1}^{R}
    t_{\mathrm{s}}^{(k,r)}.
\end{equation}
The settling region is marked by black circles in Fig.~\ref{plot:single_subject_trajectories} (top). The mean settling time across all modes and subjects is denoted by $\overline{t}_\mathrm{s}$. 
It determines the task-relevant interval as the index set
\begin{equation}
    \mathcal I_{\mathrm{end}}
    =
    \{ i \mid t_i \in [\overline t_s,\ \overline t_s + \SI{0.5}{\second}] \}.
\end{equation}
The corresponding time interval is marked as a grey area in Fig.~\ref{plot:single_subject_trajectories} (bottom).
\item[Task-relevant error (Endpoint Error)] $\poserrend$:\hfill \\
The task-relevant error is defined as the mean Euclidean distance between the mean position trajectory $\boldsymbol{\overline{p}}^{(k)}$ and the target point $\boldsymbol{p}_{\mathrm{ref}}^{(k)}$ over the task-relevant interval $\mathcal I_{\mathrm{end}}$. For each key movement $k$, the time-averaged endpoint error $e_{\mathrm{end}}^{(k)}$ is computed:
\begin{equation}
    e_{\mathrm{end}}^{(k)}
    =
    \frac{1}{|\mathcal I_{\mathrm{end}}|}
    \sum_{i\in\mathcal I_{\mathrm{end}}}
    \left\|
\overline{\bm{p}}_{i}^{(k)}
-
\bm{p}_{\mathrm{ref}}^{(k)}
\right\|_2.
\end{equation}
It is subsequently averaged across all four key movements:
\begin{equation}
    \poserrend =
    \frac{1}{K} 
    \sum_{k \in \mathcal{K}} 
    e_{\mathrm{end}}^{(k)}.
\end{equation}

% \item[Task-relevant variability (Endpoint Variance)] \hfill \\  $\posvarend$: As task-relevant variability we define the mean positional \textbf{variance} over %the task-relevant time interval  
% $\boldsymbol{t}_{\mathrm{end}}$:
% \begin{equation}
%     todo%\posvarend= \mean{\posvar_{\boldsymbol{t}_{\mathrm{end}}}}.
% \end{equation}

\item[Task-relevant variability (Endpoint Variance)] $\posvarend$:  \hfill \\ 
Using the previously defined variance trajectory $V_y(i)$, the endpoint variance was computed as the time-average over the task-relevant interval
$\mathcal I_{\mathrm{end}}$:
\begin{equation}
    \posvarend
    =
    \frac{1}{|\mathcal I_{\mathrm{end}}|}
    \sum_{i\in\mathcal I_{\mathrm{end}}}
    V_y(i).
\end{equation}

 \end{description}

\subsection{Experience measures}
%%%%%%%%%%%%%%%%%%%%%%%%%%%%%%%%%%%%%%%%%%%%%%%%%%%%%%%%%%%%%%%%%%

The following psychological measures were collected to characterize participants’ interaction experience during the task. Unless otherwise stated, all questionnaires employed a 7-point Likert scale ranging from 1 (strongly disagree) to 7 (strongly agree).

\begin{description}
    \item[Sense of Agency] Sense of agency was assessed using a 13-item questionnaire~\cite{Tapal.2017} evaluating participants’ perceived control over their movements and their outcomes. The German translation by~\cite{Bart.2023} was used. Higher overall scores indicate a stronger perceived sense of agency. The questionnaire comprises both positive and negative agency-related items.

    \item[Self-efficacy] Participants’ belief in their ability to successfully perform the task was assessed using a 5-item questionnaire~\cite{Hinds.1998}. Higher scores indicate a stronger belief in one’s capabilities. The questionnaire was translated into German.

    \item[Flow] Flow experience was assessed using the German 16-item Flow Short Scale~\cite{Rheinberg.2003}. The items capture the core flow experience, describing a mental state of deep immersion and focused attention during the task, with higher scores indicating a stronger flow experience. Two additional subscales, each consisting of three items, measure task-related worry and the perceived fit between personal capabilities and task demands. The capabilities–demands subscale uses a 9-point Likert scale, while all remaining items are rated on a 7-point Likert scale.

    \item[Usability] Usability was assessed using the Questionnaire for the Evaluation of Physically Assistive Devices (QUEAD)~\cite{Schmidtler.2017}. The questionnaire consists of 12 items and assesses perceived user experience across three subscales. \textit{Perceived usefulness} reflects the extent to which using the system is believed to improve task performance, \textit{perceived ease of use} captures the extent to which system use is perceived as effortless, and \textit{emotional response} assesses the emotional experience during the interaction. Higher scores indicate a more positive user experience. The questionnaire was translated into German.
\end{description}

All questionnaires were administered after each experimental condition.

 %%%%%%%%%%%%%%%%%%%%%%%%%%%%%%%%%%%%%%%%%%%%%%%%%%%%%%%%%%%%%%%%%%
%%%%%%%%%%%%%%%%%%%%%%%%%%%%%%%%%%%%%%%%%%%%%%%%%%%%%%%%%%%%%%%%%%
\section{Results}
%%%%%%%%%%%%%%%%%%%%%%%%%%%%%%%%%%%%%%%%%%%%%%%%%%%%%%%%%%%%%%%%%%
%%%%%%%%%%%%%%%%%%%%%%%%%%%%%%%%%%%%%%%%%%%%%%%%%%%%%%%%%%%%%%%%%%

Data from 41 participants were analyzed, of whom 17 identified as female and 24 as male. The mean participant age was 28.3 years (SD = 8.0), with a median of 26 years and an age range of 21--57 years.

Effects of automation mode as the independent variable on task and experience measures were analyzed using repeated-measures one-way ANOVAs. Greenhouse--Geisser corrections were applied when the assumption of sphericity was violated. Post-hoc pairwise comparisons with appropriate correction for multiple comparisons were conducted to assess differences between conditions. Effect sizes for pairwise comparisons are reported as Cohen’s $d_z$. Statistically significant results are marked with $^*$ for $p < .01$, and trends ($ p < .10$) are marked with $^\#$.
Statistical analyses were conducted in MATLAB~(MathWorks, Natick, MA, USA).

\begin{figure}[htbp]
    \begin{center}
    \resizebox{0.95\textwidth}{!}{%
    \input{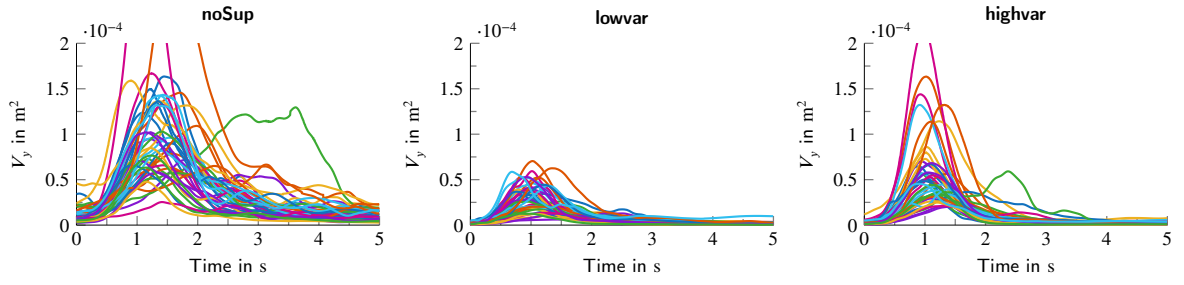}
}
    \caption{ Mean positional variance of all subjects.} 
    \label{plot:meanVarAllModes}
    \end{center}
\end{figure}

%%%%%%%%%%%%%%%%%%%%%%%%%%%%%%%%%%%%%%%%%%%%%%%%%%%%%%%%%%%%%%%%%%
\subsection{Task performance}
%%%%%%%%%%%%%%%%%%%%%%%%%%%%%%%%%%%%%%%%%%%%%%%%%%%%%%%%%%%%%%%%%%

%variance over time for all modes. significant differences in max variance for all three modes, with noSup > highVar > lowVar:

\begin{figure}[htbp]
    \begin{center}
    \input{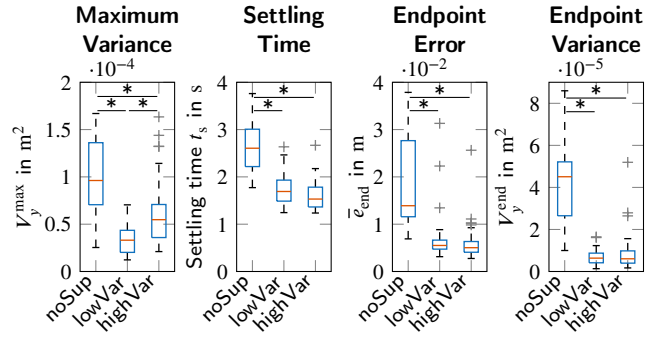}    % The printed column width is 8.4 cm.
    \caption{Maximum positional variance and task measures.} 
    \label{plot:taskMeasures}
    \end{center}
\end{figure}

Figure~\ref{plot:meanVarAllModes} illustrates the temporal evolution of positional variance for all participants, grouped by interaction modes. Based on these trajectories, the maximum positional variance was extracted for each participant and condition. The resulting values are summarized in Table~\ref{table:data}, with the corresponding distributions illustrated as boxplots in Fig.~\ref{plot:taskMeasures}. 

A qualitative inspection of the variance trajectories in Fig.~\ref{plot:meanVarAllModes} reveals substantial positional variance in the \nosup mode, along with a pronounced inter-subject spread. The \highvar mode partially reproduces this behavior: although the trajectories are slightly more constrained, many subjects still exhibit comparably high variance profiles. In contrast, the \lowvar condition shows consistently constrained variance trajectories across subjects.

Overall, this qualitative pattern aligns with the study design objectives, namely approximating the natural variance characteristics of the \nosup condition in the \highvar mode while deliberately reducing variability in the \lowvar condition. This qualitative pattern mirrors the differences quantified in the subsequent statistical analysis.

A one-way repeated-measures ANOVA revealed a significant effect of interaction mode on maximum positional variance, 
$F(1.64, 65.70) = 65.78$, $p < .001$, $\eta_p^2 = .62$ (see Table~\ref{table:data}). 
Post-hoc pairwise comparisons showed that all three modes differed significantly from one another (all $p \le .004$). 
The \nosup condition exhibited the highest maximum variance ($M = 1.11$), followed by \highvar ($M = 0.65$) and \lowvar ($M = 0.33$). 
Among the assisted modes, maximum positional variance was significantly higher in the \highvar condition than in the \lowvar condition ($p = .004$, $d_z = 0.98$), thereby confirming Hypothesis~H1 that the \highvar mode preserves more task-irrelevant variability than the \lowvar mode.

Figure~\ref{plot:taskMeasures} further summarizes the task-relevant performance measures. Significant effects of interaction mode were observed for settling time, endpoint error, and endpoint variance. 
Settling time differed significantly across modes, $F(1.33, 53.20) = 148.16$, $p < .001$, $\eta_p^2 = .79$, with the longest times in \nosup ($M = 2.64$), followed by \lowvar ($M = 1.72$) and \highvar ($M = 1.61$). 
Similarly, endpoint error showed a significant effect of interaction mode, $F(1.05, 41.88) = 35.08$, $p < .001$, $\eta_p^2 = .47$, with highest error in \nosup ($M = 2.35$) compared to \lowvar ($M = 0.69$) and \highvar ($M = 0.60$). 
Endpoint variance also differed significantly between modes, $F(1.17, 46.61) = 79.10$, $p < .001$, $\eta_p^2 = .66$, with the highest values in \nosup ($M = 4.51$), compared to \lowvar ($M = 0.69$) and \highvar ($M = 0.87$).

Post-hoc comparisons confirmed that both assisted conditions (\lowvar and \highvar) significantly outperformed \nosup across all three measures (all $p < .001$), thereby confirming Hypothesis~H2a. 
Importantly, no significant differences were observed between \highvar and \lowvar for settling time ($p = .409$, $d_z = 0.36$), endpoint error ($p = .949$, $d_z = 0.20$), or endpoint variance ($p = .866$, $d_z = -0.21$), confirming Hypothesis~H2b that preserving task-irrelevant variability does not compromise task-relevant performance.

\begin{table*}[bp]
    \caption{ Mean $\pm$ standard deviation of task measures (top), results of repeated-measures ANOVA (middle) and pairwise comparisons (bottom). Statistical significances are marked with  $^*$ for $p<0.01$ and trends with $^\#$ for $p<0.1$.}
    \label{table:data}
    \small
    \centering
    \setlength{\tabcolsep}{6pt} % <-- default is 6pt
    %\begin{tabular}{ccccccccccccc}
    \begin{tabular*}{\tblwidth}{@{\extracolsep{\fill}}ccccc}
        \toprule %\hline
        & \textbf{\begin{tabular}[c]{@{}c@{}}Max. Var. \textnormal{in} $10^{-4}\,$\SI{}{\meter\squared} \end{tabular}}  
        & \textbf{\begin{tabular}[c]{@{}c@{}}Settling Time \textnormal{in} \SI{}{\second}\end{tabular}} 
        & \textbf{\begin{tabular}[c]{@{}c@{}}Endpoint Error \textnormal{in} $10^{-2}\,$\SI{}{\meter}\end{tabular}} 
        & \textbf{\begin{tabular}[c]{@{}c@{}}Endpoint Var. \textnormal{in} $10^{-5}\,$\SI{}{\meter\squared}\end{tabular}} 
        \\
       \midrule
        noSup & \num{1.11 \pm 0.61} & \num{2.64 \pm 0.52} & \num{2.35 \pm 2.13}  & \num{4.51\pm 2.68}   \\
        lowVar & \num{0.33\pm 0.16} &  \num{1.72 \pm 0.16} & \num{0.69\pm 0.50}  &\num{0.69\pm 0.35}   \\
        highVar & \num{0.65\pm 0.43} &  \num{1.61\pm 0.31}& \num{0.60 \pm 0.38} & \num{0.87\pm 0.88} \\
        \midrule
        df1,df2 & 1.64,65.70 & 1.33,53.20 & 1.05,41.88 & 1.17,46.61  \\
        
        F(df1,df2) & 65.78 & 148.16 & 35.08 & 79.10  \\

        $\eta_p^2$ & 0.62 & 0.79 & 0.47 & 0.66  \\

        $p_{\mathrm{ANOVA}}$ & $<0.001^{*}$ & $<0.001^{*}$ & $<0.001^{*}$ & $<0.001^{*}$  \\
        \midrule
        noSup - lowVar & $<0.001^{*}$ & $<0.001^{*}$ & $<0.001^{*}$ & $<0.001^{*}$ \\
        Cohen's $d_z$ & 1.49 & 2.04 & 0.97 & 1.51  \\
        noSup - highVar & $<0.001^{*}$ & $<0.001^{*}$ & $<0.001^{*}$ & $<0.001^{*}$ \\
         Cohen's $d_z$ & 1.10 & 1.98 &  0.90 & 1.34  \\
        lowVar - highVar & $0.004^{*}$ & 0.409 & 0.949 & 0.866 \\
         Cohen's $d_z$ & -0.90 & 0.47 & 0.23 & -0.21  \\
        \bottomrule %\hline
    \end{tabular*}
\end{table*}

\begin{table*}[bp]
    \caption{Mean $\pm$ standard deviation of interaction experience measures (top), results of repeated-measures ANOVA (middle) and pairwise comparisons (bottom). Statistical significances are marked with $^*$ for $p<0.01$ and trends with $^\#$ for $p<0.1$.}
    \label{table:experience}
    \small
    \centering
    \setlength{\tabcolsep}{6pt}
    \begin{tabular*}{\tblwidth}{@{\extracolsep{\fill}}ccccccccc}
        \toprule
        & \textbf{\begin{tabular}[c]{@{}c@{}}Sense of \\ Agency\end{tabular}}
        & \textbf{Self-efficacy}
        & \textbf{Flow}
        & \textbf{Usability}
        & \textbf{\begin{tabular}[c]{@{}c@{}}Perceived \\ Usefulness\end{tabular}}
        & \textbf{\begin{tabular}[c]{@{}c@{}}Perceived \\ Ease of Use\end{tabular}}
        & \textbf{\begin{tabular}[c]{@{}c@{}}Emotional \\ Response\end{tabular}} \\
        \midrule

        noSup 
        & \num{4.82 \pm 0.91}
        & \num{5.04 \pm 0.81}
        & \num{4.55 \pm 0.88}
        & \num{3.80 \pm 0.87}
        & \num{3.80 \pm 1.02}
        & \num{3.69 \pm 0.98}
        & \num{3.99 \pm 1.10} \\

        lowVar 
        & \num{5.35 \pm 0.82}
        & \num{5.96 \pm 0.65}
        & \num{5.16 \pm 0.84}
        & \num{5.27 \pm 0.84}
        & \num{5.60 \pm 1.00}
        & \num{4.92 \pm 1.00}
        & \num{5.45 \pm 1.12} \\

        highVar 
        & \num{5.46 \pm 0.64}
        & \num{6.14 \pm 0.60}
        & \num{5.29 \pm 0.88}
        & \num{5.85 \pm 0.69}
        & \num{5.86 \pm 0.82}
        & \num{5.80 \pm 0.76}
        & \num{5.93 \pm 0.86} \\
        \midrule

        df1 
        & 1.33 & 1.70 & 1.71 & 2 & 2 & 2 & 2 \\

        df2 
        & 53.06 & 68.16 & 68.54 & 80 & 80 & 80 & 80 \\

        F(df1,df2) 
        & 9.31 & 57.50 & 18.41 & 82.94 & 64.66 & 61.71 & 47.79 \\

        $\eta_p^2$ 
        & 0.19 & 0.59 & 0.31 & 0.67 & 0.62 & 0.61 & 0.54 \\

        $p_{\mathrm{ANOVA}}$ 
        & $0.0016^{*}$ & $<0.001^{*}$ & $<0.001^{*}$ & $<0.001^{*}$ & $<0.001^{*}$ & $<0.001^{*}$ & $<0.001^{*}$ \\
        \midrule

        noSup - lowVar 
        & $0.0094^{*}$ & $<0.001^{*}$ & $0.0057^{*}$ & $<0.001^{*}$ & $<0.001^{*}$ & $<0.001^{*}$ & $<0.001^{*}$ \\

        Cohen's $d_z$ 
        & -0.41 & -1.30 & -0.30 & -1.37 & -1.31 & -0.97 & -1.08 \\

        noSup - highVar 
        & $0.001^{*}$ & $<0.001^{*}$ & $<0.001^{*}$ & $<0.001^{*}$ & $<0.001^{*}$ & $<0.001^{*}$ & $<0.001^{*}$ \\

        Cohen's $d_z$ 
        & -0.64 & -1.34 & -0.75 & -1.83 & -1.60 & -1.70 & -1.44 \\

        lowVar - highVar 
        & 0.805 & 0.467 & 0.793 & $0.004^{*}$ & 0.377 & $<0.001^{*}$ & $0.087^{\#}$ \\

        Cohen's $d_z$ 
        & -0.17 & -0.32 & -0.17 & -0.61 & -0.25 & -0.76 & -0.38 \\

        \bottomrule
    \end{tabular*}
\end{table*}

%%%%%%%%%%%%%%%%%%%%%%%%%%%%%%%%%%%%%%%%%%%%%%%%%%%%%%%%%%%%%%%%%%
\subsection{Human interaction experience}
%%%%%%%%%%%%%%%%%%%%%%%%%%%%%%%%%%%%%%%%%%%%%%%%%%%%%%%%%%%%%%%%%%

\begin{figure}[htbp]
    \begin{center}
    \resizebox{\textwidth}{!}{%
    \input{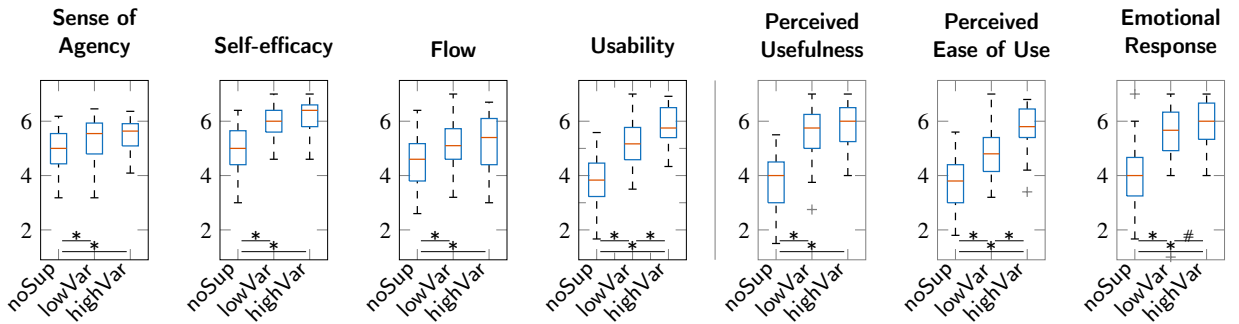}   % The printed column width is 8.4 cm.
    }
    \caption{Interaction experience measures, showing the overall experience measures (left) and the usability subscales (right).} 
    \label{plot:experienceMeasures}
    \end{center}
\end{figure}

Figure~\ref{plot:experienceMeasures} (left) summarizes the self-reported experience measures, including sense of agency, self-efficacy, flow, and usability. 
Numerical results are reported in Table~\ref{table:data}.
Significant effects of interaction mode were also observed for all four experience measures. 
Sense of agency differed significantly across modes, $F(1.33, 53.06) = 9.31$, $p = .0016$, $\eta_p^2 = .19$, with higher ratings in \lowvar ($M = 5.35$) and \highvar ($M = 5.46$) compared to \nosup ($M = 4.82$). 
Self-efficacy showed a strong effect of interaction mode, $F(1.70, 68.16) = 57.50$, $p < .001$, $\eta_p^2 = .59$, with both assisted modes (\lowvar: $M = 5.96$, \highvar: $M = 6.14$) exceeding \nosup ($M = 5.04$). 
Flow also differed significantly between modes, $F(1.71, 68.54) = 18.41$, $p < .001$, $\eta_p^2 = .31$, again with higher ratings under assistance (\lowvar: $M = 5.16$, \highvar: $M = 5.29$) compared to \nosup ($M = 4.55$). 
Finally, usability revealed a strong effect of interaction mode, $F(2, 80) = 82.94$, $p < .001$, $\eta_p^2 = .67$, with the highest ratings observed in \highvar ($M = 5.85$), followed by \lowvar ($M = 5.27$) and \nosup ($M = 3.80$).

Post-hoc comparisons confirmed that both assisted modes were rated significantly higher than \nosup across all experience measures (all $p < .01$). 
Importantly, usability was significantly higher in the \highvar condition than in the \lowvar condition ($p = .004$, $d_z = 0.76$), while no significant differences between the assisted modes were observed for sense of agency, self-efficacy, or flow. 
Taken together, these findings partly confirm Hypothesis~H3, indicating that variability-aware control enhances interaction experience relative to no assistance and selectively improves perceived usability compared to variability-constraining assistance.

Given that usability was the only experience measure showing a significant difference between the \highvar and \lowvar conditions, we next examined the subscales of the usability questionnaire to identify which components contributed to this effect. The distributions of the usability subscales are shown in Figure~\ref{plot:experienceMeasures} (right).

A similar pattern was observed for the usability subscales. 
Perceived usefulness differed significantly across modes, $F(2, 80) = 64.66$, $p < .001$, $\eta_p^2 = .62$, with higher ratings in both assisted conditions (\lowvar: $M = 5.60$, \highvar: $M = 5.86$) compared to \nosup ($M = 3.80$). However, no significant difference was observed between \highvar and \lowvar ($p = .377$, $d_z = 0.31$). 

Perceived ease of use also showed a significant effect of interaction mode, $F(2, 80) = 61.71$, $p < .001$, $\eta_p^2 = .61$, with ratings increasing from \nosup ($M = 3.69$) to \lowvar ($M = 4.92$) and further to \highvar ($M = 5.80$). Importantly, perceived ease of use was significantly higher in the \highvar condition than in the \lowvar condition ($p < .001$, $d_z = 1.00$). 

Emotional response likewise differed significantly between modes, $F(2, 80) = 47.79$, $p < .001$, $\eta_p^2 = .54$, with higher ratings under assistance (\lowvar: $M = 5.45$, \highvar: $M = 5.93$) compared to \nosup ($M = 3.99$). The difference between \highvar and \lowvar did not reach conventional statistical significance, but showed a trend toward higher ratings in the \highvar condition ($p = .087$, $d_z = 0.49$).

\FloatBarrier

 \section{Discussion}
The aim of the present study was to test whether varia\-bility-aware control in a shared control setting enables an increased interaction experience compared to a conventional controller design, while at the same time resulting in similar task performance. 
The results demonstrate that variability-aware control achieves comparable task performance to conventional shared control while enhancing perceived usability. Specifically, the variability-respecting controller increased task-irrelevant variability (maximum positional variance) without compromising task-relevant performance measures. At the same time, both assisted modes improved interaction experience relative to unsupported interaction, with variability-aware control yielding the highest usability ratings. Together, these findings suggest that preserving natural movement variability can enhance interaction quality without sacrificing objective performance. %Die Zusammenfassung habe ich jetzt sehr kurz gehalten. 

\subsection{Preserving natural movement variability}
%%%%%%%%%%%%%%%%%%%%%%%%%%%%%%%%%%%%%%%%%%%%%%%%%%%%%%%%%%%%%%%%%%

%maximum variability: 
In line with Hypothesis~H1, we observed a significant increase in task-irrelevant variability (maximum positional variance) in the \highvar mode compared to the \lowvar mode. This validates the effectiveness of the variability-respecting controller HVROC that was previously only conceptually introduced, demonstrating that it preserves a more natural variability than a standard controller. However, a reproduction of the same level of task-irrelevant variability as observed in the \textit{noSup} condition was not achieved. 

This discrepancy may be attributed to the fact that any form of physical assistance introduces stabilizing dynamics into the human-machine system, thereby reducing overall movement variance relative to unsupported execution. 
Moreover, because human movement variability partly arises from signal-dependent motor noise, reducing the human’s required control input through assistance inherently attenuates this source of variability.
In addition, the \highvar parametrization deliberately included a damping component to ensure comparable performance between assisted modes, further constraining the achievable variability level. %hier auch auf potential future work verweisen? (passiert bereits im outlook, deshalb hab ich es hier nicht nochmal erwähnt) 

Importantly, preserving variability does not imply reproducing unsupported motor noise, but rather maintaining the statistical structure of variability in dimensions that are functionally redundant for task success. This selective preservation may be sufficient to sustain the internal prediction structures underlying agency, even if absolute variance levels remain lower than in unsupported movement.

%task-relevant measures:
Nevertheless, the task-relevant measures settling time, endpoint error, and endpoint variance revealed no significant differences between the \highvar and \lowvar modes, but a significant performance increase compared to \nosup. This clearly shows the effectiveness of the control design regarding task performance, confirming~Hypothesis H2a. Furthermore, it allows us to confirm Hypothesis~H2b, stating that no significant difference in task-relevant performance could be observed between variability-respecting and variability-constraining control. This finding is important as it highlights the observation that preserving natural variability patterns does not come at the cost of task performance. Critically, the same level of task performance between \highvar and \lowvar allows us to attribute differences in user experience specifically to the variability manipulation, without confounding effects from performance disparities. 

%eher nicht, da nicht signifikant: (discuss why highvar tends to have even better performance than lowvar: maybe because preserving natural variability patterns allows for more intuitive and efficient movements, leading to better overall control and task execution. Future studies with larger sample sizes could further investigate these trends to determine if they reach statistical significance.)

%%%%%%%%%%%%%%%%%%%%%%%%%%%%%%%%%%%%%%%%%%%%%%%%%%%%%%%%%%%%%%%%%%
\subsection{Human interaction experience}
%%%%%%%%%%%%%%%%%%%%%%%%%%%%%%%%%%%%%%%%%%%%%%%%%%%%%%%%%%%%%%%%%%

Although task performance was comparable between \lowvar and \highvar, perceived usability differed significantly across all three modes, with \highvar rated highest, followed by \lowvar and \nosup. Accordingly, this finding partly confirms Hypothesis~H3 regarding perceived usability, demonstrating a positive impact of variability-aware control concerning that experience measure. This pattern indicates that preserving natural variability enhances perceived usability beyond performance-related effects. Because task outcomes were equivalent between assisted modes, the observed usability differences cannot be attributed to differences in effectiveness, but rather to qualitative differences in how assistance was experienced.

A more detailed analysis of the usability subscales further supports this interpretation. Perceived ease of use was significantly higher in the \highvar condition than in the \lowvar condition, while emotional response showed a positive trend in the same direction. In contrast, no differences were observed in perceived usefulness, which closely mirrored the comparable task performance across assisted modes. This dissociation suggests that users primarily evaluate usefulness based on task success, whereas ease of use and affective response are shaped by the manner in which assistance is rendered.

From a theoretical perspective, preserving task-irrelevant variability may maintain the natural sensorimotor contingencies that underlie users’ internal movement expectations. Even when task outcomes are identical, constraining variability may subtly alter the felt dynamics of interaction, leading to a reduced sense of intuitiveness or effortlessness. By contrast, allowing natural variability may better align with internal motor models, thereby promoting a more comfortable interaction experience. This interpretation aligns with distinctions between pragmatic effectiveness and experiential or hedonic qualities of interaction~\cite{Hassenzahl.2008}.

Turning to sense of agency, a different pattern emerges. Although sense of agency ratings increased under assisted conditions relative to \nosup, no significant differences were observed between \highvar and \lowvar. At first glance, the increase in agency under assistance may appear counterintuitive, as greater automation involvement might be expected to reduce the feeling of being responsible for action outcomes. However, this result aligns with motor prediction models, which posit that agency is strengthened when predicted and actual sensory consequences match~\cite{Wen.2015}. Because both assisted modes substantially improved task success, they likely enhanced the congruence between intended and achieved outcomes, thereby increasing perceived agency.

The absence of differences between \highvar and \lowvar can be understood in light of cue-integration accounts of agency~\cite{Haggard.2012, Synofzik.2013}. Agency judgments are shaped by multiple cues, including low-level sensorimotor prediction mechanisms and higher-level outcome-related information. In the present task, strong performance-related cues may have dominated agency judgments, overshadowing more subtle differences in task-irrelevant variability structure. Moreover, the variability manipulation was confined to dimensions functionally redundant for task success. While such manipulations may influence the qualitative feel of movement, they may not have been sufficiently salient to alter explicit agency ratings. Together, these findings suggest that agency in this context was primarily driven by successful goal attainment rather than by fine-grained differences in variability structure.

A similar performance-driven pattern was observed for flow and self-efficacy. Both constructs were significantly higher in the assisted modes compared to \nosup, with no differences between \highvar and \lowvar. Although conceptually distinct, both are closely related to perceived task competence and mastery. Flow theory posits that optimal experience emerges when task demands are well matched to perceived skill level~\cite{Csikszentmihalyi.1990, Csikszentmihalyi.1989}. By reducing performance demands and increasing task success, both assisted modes likely facilitated a favorable challenge–skill balance, thereby enhancing flow.

Similarly, self-efficacy reflects individuals’ beliefs about their capability to successfully execute task actions and is strongly shaped by mastery experiences~\cite{Bandura.1997}. Because both assisted modes enabled comparable improvements in task-relevant performance, participants likely experienced similar levels of competence and control in \highvar and \lowvar. Importantly, although automation contributed to task success, participants may still have perceived themselves as the primary agents responsible for performance, particularly in a physically coupled setting where active motor engagement was maintained~\cite{Berberian.2019}. Consequently, flow and self-efficacy appear to have been predominantly driven by perceived task mastery rather than by qualitative differences in movement variability.

Taken together, these results indicate that different experiential constructs respond to distinct aspects of control design. While agency, flow, and self-efficacy were largely shaped by successful task performance, perceived usability was sensitive to how assistance was physically instantiated. In this context, allowing natural variability likely better matched participants’ internal motor models---and their expectations of how movement should feel---thereby promoting a more intuitive and comfortable interaction experience even when objective performance was comparable.

\subsection{Implications for pHMI design}
%%%%%%%%%%%%%%%%%%%%%%%%%%%%%%%%%%%%%%%%%%%%%%%%%%%%%%%%%%%%%%%%%%

The findings of the present study have important implications for the design of pHMI systems. Beyond demonstrating that variability-aware control can enhance interaction experience, the results highlight that interaction quality can be systematically shaped without compromising task performance. In particular, the observed improvements in perceived ease of use and emotional response indicate that the manner in which assistance is rendered plays a critical role in how supportive and intuitive a system is perceived to be.

From a control design perspective, the equivalence in task performance between variability-respecting and varia\-bility-constraining assistance suggests that preserving natural human movement variability does not inherently compromise efficiency or accuracy. Instead, variability-aware control provides a means to shape the qualitative characteristics of the interaction, such as perceived ease of use and affective response, without altering objective task outcomes. This decoupling of performance and experience opens a design space in which multiple control strategies may be functionally equivalent but experientially distinct.

The observed improvements in perceived ease of use and emotional response further suggest that respecting natural variability may support more intuitive and pleasant interactions. These experiential aspects are particularly relevant in pHMI scenarios where users remain physically engaged with the system and continuously adapt their movements, as they influence how supportive, transparent, and comfortable assistance feels during interaction.

Importantly, the absence of differences between varia\-bility-aware and variability-constraining control in perceived usefulness indicates that users primarily evaluate usefulness based on task success, while other experiential qualities are shaped by the manner in which assistance is delivered. This distinction underscores the importance of considering interaction experience as an explicit design objective in pHMI systems, especially in contexts where task performance is already reliably supported by automation.

Finally, while the present study focuses on short-term interactions, the enhanced ease of use and positive emotional response associated with variability-aware control point toward potential benefits for long-term use and acceptance. This may be particularly relevant for applications such as rehabilitation or assistive robotics, where sustained interaction quality is critical.

Overall, the results demonstrate that preserving structured movement variability improves interaction quality without sacrificing task performance. Accordingly, variability-aware principles should be explicitly incorporated into control design for tightly coupled pHMI systems to achieve both effectiveness and experiential quality.

%%%%%%%%%%%%%%%%%%%%%%%%%%%%%%%%%%%%%%%%%%%%%%%%%%%%%%%%%%%%%%%%%%
\subsection{Limitations and future research directions}
%%%%%%%%%%%%%%%%%%%%%%%%%%%%%%%%%%%%%%%%%%%%%%%%%%%%%%%%%%%%%%%%%%

Although the study was conducted with a comparatively large participant pool that included participants from two different universities, some limitations remain that motivate future studies investigating variability-aware control.
%Task specificity and generalizability: 
First, the study focused on an isolated point-to-point movement task to enable controlled manipulation of movement variability. Therefore, it remains an open question to what extent the observed benefits of variability-aware control generalize to more complex pHMI tasks. Future studies could extend variability-aware control to more complex pHMI tasks, such as tasks with higher redundancy, sequential subtasks, or dynamically changing goals, thereby providing insights into how these principles generalize to realistic interaction scenarios.
Nevertheless, this is the first study that investigated the effects of variability-aware control on human experience. Therefore, it was important to choose a design that allows for a clear interpretation of variability-related effects, which we achieved with this rather isolated point-to-point movement task.

%Extent of preserved variability:
Although the \highvar condition preserved significantly more task-irrelevant variability than the \lowvar condition, it did not reproduce the same level of variability observed in the \nosup condition. This results from the deliberate parametrization to ensure comparable task performance across the assisted modes, which we achieved. 
However, for future work it would be interesting to investigate stronger or adaptive variability preservation strategies to better understand how different degrees of variability influence both task performance and interaction experience, and whether an optimal balance exists between support and flexibility.

%Uniform control parametrization across participants:
Finally, all participants interacted with identical control laws and parameter settings. While this supports comparability across conditions, it does not account for individual differences in movement variability, motor strategies, or preferences, which may influence how variability-aware assistance is perceived. Future research should address individual differences in movement variability and user preferences. Personalizing variability-aware control strategies to individual users, for example by adapting controller parameters to subject-specific variability profiles, may further enhance interaction quality and could be particularly relevant for assistive or rehabilitative pHMI applications. 

%Short-term interaction scope: 
%The study investigated immediate effects of variability manipulation within a single-session interaction. Potential longer-term effects on learning, adaptation, trust, or sustained user experience were not assessed.

%%%%%%%%%%%%%%%%%%%%%%%%%%%%%%%%%%%%%%%%%%%%%%%%%%%%%%%%%%%%%%%%%%
%\subsection{Future Research Directions}

%The present findings motivate several directions for future research. First, while the variability-aware controller preserved task-irrelevant variability to a greater extent than the variability-constraining mode, the level of preserved variability was deliberately limited to ensure comparable task performance. 

%Second, the current study focused on a controlled point-to-point task to isolate the effects of variability manipulation. 

%Third, the absence of significant differences between variability-aware and variability-constraining control for measures such as sense of agency, flow, and self-efficacy suggests that these experiential dimensions may evolve over longer interaction periods. Longitudinal studies could examine whether preserving natural movement variability influences learning, adaptation, trust, or sustained engagement with the system over time.

%game-theoretic control ansprechen?

 \section{Conclusions}

pHMI is characterized by tight physical coupling, where control policies are directly experienced through haptic feedback. While conventional shared-control strategies prioritize task performance, they largely ignore a fundamental characteristic of human motor behavior: structured movement variability. In particular, natural movements exhibit low variability in task-relevant dimensions and higher variability in task-irrelevant dimensions. Suppressing this structure may alter sensorimotor contingencies and affect interaction experience.

In this study, we experimentally investigated whether explicitly preserving task-irrelevant variability influences human experience without compromising task performance. Using a human-variability-respecting optimal controller, we manipulated variability levels in a controlled haptic point-to-point task under matched task-relevant performance. A variability-constraining mode (\lowvar), a variability-respecting mode (\highvar), and a no-support (\nosup) baseline were compared in a within-subject design.

Results from 41 participants demonstrate that variability-aware control significantly increases task-irrelevant variability relative to conventional control while maintaining equivalent task performance. Crucially, usability ratings were significantly higher under variability-respecting control, primarily driven by improved perceived ease of use. Other experiential constructs (agency, flow, self-efficacy) were primarily performance-driven and did not differ between assisted modes.

These findings establish a causal link between variability structure and interaction experience in tightly coupled pHMI. They show that control policies can be experientially distinct despite equivalent performance, and that preserving natural movement variability enhances perceived usability. 
Variability-aware control should therefore be regarded not as an optional refinement, but as a deliberate design principle for human-centered shared-control architectures that aim to combine performance with sustained interaction quality.

% Numbered list
% Use the style of numbering in square brackets.
% If nothing is used, default style will be taken.
%\begin{enumerate}[a)]
%\item 
%\item 
%\item 
%\end{enumerate}  

% Unnumbered list
%\begin{itemize}
%\item 
%\item 
%\item 
%\end{itemize}  

% Description list
%\begin{description}
%\item[]
%\item[] 
%\item[] 
%\end{description}  

% % Figure
% \begin{figure}%[]
%   \centering
% %    \includegraphics{}
%     \caption{}\label{fig1}
% \end{figure}

% \begin{table}%[]
% \caption{}\label{tbl1}
% \begin{tabular*}{\tblwidth}{@{}LL@{}}
% \toprule
%   &  \\ % Table header row
% \midrule
%  & \\
%  & \\
%  & \\
%  & \\
% \bottomrule
% \end{tabular*}
% \end{table}

% Uncomment and use as the case may be
%\begin{theorem} 
%\end{theorem}

% Uncomment and use as the case may be
%\begin{lemma} 
%\end{lemma}

%% The Appendices part is started with the command \appendix;
%% appendix sections are then done as normal sections
\appendix

\section*{Funding}
This research did not receive any specific grant from funding agencies in the public, commercial, or not-for-profit sectors.

\section*{Data availability}
The research data will be made available on request.  

\section*{Declaration of generative AI and AI-assisted technologies in the manuscript preparation process}%\label{}
During the preparation of this work the authors used ChatGPT (OpenAI) for language editing and text refinement. All scientific content and interpretations were developed by the authors. After using this tool, the authors reviewed and edited the content as needed and take full responsibility for the content of the published article.

% To print the credit authorship contribution details
%\printcredits   

%% Loading bibliography style file
%\bibliographystyle{model1-num-names}
\bibliographystyle{cas-model2-names}

% Loading bibliography database
\bibliography{1_tex/bibliography}

@article{Abend.1982,
  title = {Human Arm Trajectory Formation},
  author = {Abend, W. and Bizzi, E. and Morasso, P.},
  year = 1982,
  journal = {Brain},
  number = {105},
  doi = {10.1093/brain/105.2.331}
}

@book{Bandura.1997,
  title = {Self-Efficacy: {{The}} Exercise of Control},
  shorttitle = {Self-Efficacy},
  author = {Bandura, Albert},
  year = 1997,
  volume = {11},
  publisher = {Freeman}
}

@article{Bart.2023,
  title = {A {{German}} Translation and Validation of the Sense of Agency Scale},
  author = {Bart, Victoria K. E. and Wenke, Dorit and Rieger, Martina},
  year = 2023,
  journal = {Frontiers in Psychology},
  volume = {14},
  publisher = {Frontiers},
  doi = {10.3389/fpsyg.2023.1199648},
  urldate = {2025-04-02},
  langid = {english}
}

@inproceedings{Bennett.2023,
  title = {How Does {{HCI Understand Human Agency}} and {{Autonomy}}?},
  booktitle = {Proceedings of the 2023 {{CHI Conference}} on {{Human Factors}} in {{Computing Systems}}},
  author = {Bennett, Dan and Metatla, Oussama and Roudaut, Anne and Mekler, Elisa D.},
  year = 2023,
  series = {{{CHI}} '23},
  publisher = {Association for Computing Machinery},
  address = {New York, NY, USA},
  doi = {10.1145/3544548.3580651},
  urldate = {2026-02-12},
  isbn = {978-1-4503-9421-5}
}

@article{Berberian.2019,
  title = {Man-Machine Teaming: A Problem of Agency},
  author = {Berberian, B.},
  year = 2019,
  journal = {IFAC-PapersOnLine},
  volume = {51},
  number = {34},
  doi = {10.1016/j.ifacol.2019.01.049}
}

@incollection{Braun.2023,
  title = {Using a {{Collaborative Robotic Arm}} as {{Human-Machine Interface}}: {{System Setup}} and {{Application}} to {{Pose Control Tasks}}},
  booktitle = {2023 {{IEEE International Conference}} on {{Robotics}} and {{Automation}} ({{ICRA}})},
  author = {Braun, Christian A. and Haide, Ludwig and Fischer, Lars and Kille, Sean and Varga, Balint and Rothfuss, Simon and Hohmann, S{\"o}ren},
  year = 2023,
  publisher = {IEEE},
  doi = {10.1109/ICRA48891.2023.10161348},
  isbn = {979-8-3503-2365-8}
}

@inproceedings{Christen.2019,
  title = {Demonstration-{{Guided Deep Reinforcement Learning}} of {{Control Policies}} for {{Dexterous Human-Robot Interaction}}},
  booktitle = {2019 {{International Conference}} on {{Robotics}} and {{Automation}} ({{ICRA}})},
  author = {Christen, Sammy and Stev{\v s}i{\'c}, Stefan and Hilliges, Otmar},
  year = 2019,
  doi = {10.1109/ICRA.2019.8794065},
  urldate = {2026-02-25}
}

@book{Csikszentmihalhi.1997,
  title = {Finding {{Flow}}: {{The Psychology}} of {{Engagement}} with {{Everyday Life}}},
  author = {Csikszentmihalhi, Mihaly},
  year = 1997,
  publisher = {Basic Books},
  address = {New York}
}

@article{Csikszentmihalyi.1989,
  title = {Optimal Experience in Work and Leisure},
  author = {Csikszentmihalyi, Mihaly and LeFevre, Judith},
  year = 1989,
  journal = {Journal of Personality and Social Psychology},
  volume = {56},
  number = {5},
  doi = {10.1037/0022-3514.56.5.815}
}

@article{Csikszentmihalyi.1990,
  title = {{{FLOW}}: {{The Psychology}} of {{Optimal Experience}}},
  author = {Csikszentmihalyi, Mihaly},
  year = 1990,
  langid = {english}
}

@article{Dong.2020,
  title = {Physical Human--Robot Interaction Force Control Method Based on Adaptive Variable Impedance},
  author = {Dong, Jianwei and Xu, Jianming and Zhou, Qiaoqian and Hu, Songda},
  year = 2020,
  journal = {Journ. of the Franklin Institute},
  volume = {357},
  number = {12},
  doi = {10.1016/j.jfranklin.2020.06.007}
}

@misc{Eloundou.2023,
  title = {{{GPTs}} Are {{GPTs}}: {{An Early Look}} at the {{Labor Market Impact Potential}} of {{Large Language Models}}},
  shorttitle = {{{GPTs}} Are {{GPTs}}},
  author = {Eloundou, Tyna and Manning, Sam and Mishkin, Pamela and Rock, Daniel},
  year = 2023,
  number = {arXiv:2303.10130},
  eprint = {2303.10130},
  primaryclass = {econ},
  publisher = {arXiv},
  doi = {10.48550/arXiv.2303.10130},
  urldate = {2025-03-21},
  archiveprefix = {arXiv},
  langid = {english}
}

@article{Engelbrecht.2001,
  title = {Minimum {{Principles}} in {{Motor Control}}},
  author = {Engelbrecht, Sascha E.},
  year = 2001,
  journal = {Journal of Mathematical Psychology},
  volume = {45},
  number = {3},
  publisher = {Elsevier BV},
  doi = {10.1006/jmps.2000.1295},
  pmid = {11401453}
}

@article{Farooq.2016a,
  title = {Human-Computer Integration},
  author = {Farooq, Umer and Grudin, Jonathan},
  year = 2016,
  journal = {Interactions},
  volume = {23},
  number = {6},
  doi = {10.1145/3001896},
  urldate = {2024-09-18},
  langid = {english}
}

@article{FirminodeSouza.2025,
  title = {Trust and {{Trustworthiness}} from {{Human-Centered Perspective}} in {{Human}}--{{Robot Interaction}} ({{HRI}})---{{A Systematic Literature Review}}},
  author = {{Firmino de Souza}, Debora and Sousa, Sonia and {Kristjuhan-Ling}, Kadri and Dunajeva, Olga and Roosileht, Mare and Pentel, Avar and M{\~o}ttus, Mati and Can {\"O}zdemir, Mustafa and Grat{\v s}jova, {\v Z}anna},
  year = 2025,
  journal = {Electronics},
  volume = {14},
  number = {8},
  publisher = {Multidisciplinary Digital Publishing Institute},
  doi = {10.3390/electronics14081557},
  urldate = {2026-02-12},
  copyright = {http://creativecommons.org/licenses/by/3.0/},
  langid = {english}
}

@article{Fitzsimons.2020,
  title = {Task-Based Hybrid Shared Control for Training through Forceful Interaction},
  author = {Fitzsimons, Kathleen and Kalinowska, Aleksandra and Dewald, Julius P and Murphey, Todd D},
  year = 2020,
  journal = {The International Journal of Robotics Research},
  volume = {39},
  number = {9},
  publisher = {SAGE Publications Ltd STM},
  doi = {10.1177/0278364920933654},
  urldate = {2025-08-20},
  langid = {english}
}

@article{Fitzsimons.2022,
  title = {Ergodic {{Shared Control}}: {{Closing}} the {{Loop}} on {{pHRI Based}} on {{Information Encoded}} in {{Motion}}},
  shorttitle = {Ergodic {{Shared Control}}},
  author = {Fitzsimons, Kathleen and Murphey, Todd D.},
  year = 2022,
  journal = {J. Hum.-Robot Interact.},
  volume = {11},
  number = {4},
  doi = {10.1145/3526106},
  urldate = {2025-08-20}
}

@article{Flad.2017,
  title = {Cooperative Shared Control Driver Assistance Systems Based on Motion Primitives and Differential Games},
  author = {Flad, Michael and Frohlich, Lukas and Hohmann, Soren},
  year = 2017,
  journal = {IEEE Transactions on Human-Machine Systems},
  volume = {47},
  number = {5},
  publisher = {{Institute of Electrical and Electronics Engineers (IEEE)}},
  doi = {10.1109/thms.2017.2700435}
}

@inproceedings{Gribovskaya.2011,
  title = {Motion Learning and Adaptive Impedance for Robot Control during Physical Interaction with Humans},
  booktitle = {2011 {{IEEE International Conference}} on {{Robotics}} and {{Automation}}},
  author = {Gribovskaya, Elena and Kheddar, Abderrahmane and Billard, Aude},
  year = 2011,
  publisher = {IEEE},
  address = {Shanghai, China},
  doi = {10.1109/ICRA.2011.5980070},
  urldate = {2025-03-20},
  isbn = {978-1-61284-386-5},
  langid = {english}
}

@article{Haggard.2012,
  title = {Sense of Agency},
  author = {Haggard, Patrick and Chambon, Valerian},
  year = 2012,
  journal = {Current Biology},
  volume = {22},
  number = {10},
  doi = {10.1016/j.cub.2012.02.040},
  pmid = {22625851}
}

@article{Harris.1998,
  title = {Signal-Dependent Noise Determines Motor Planning},
  author = {Harris, C. M. and Wolpert, D. M.},
  year = 1998,
  journal = {Nature},
  volume = {394},
  number = {6695},
  publisher = {Nature Publishing Group},
  doi = {10.1038/29528},
  pmid = {9723616}
}

@article{Hassenzahl.2008,
  title = {How Motivational Orientation Influences the Evaluation and Choice of Hedonic and Pragmatic Interactive Products: {{The}} Role of Regulatory Focus},
  shorttitle = {How Motivational Orientation Influences the Evaluation and Choice of Hedonic and Pragmatic Interactive Products},
  author = {Hassenzahl, Marc and Sch{\"o}bel, Markus and Trautmann, Tibor},
  year = 2008,
  journal = {Interacting with Computers},
  volume = {20},
  number = {4-5},
  doi = {10.1016/j.intcom.2008.05.001},
  urldate = {2026-02-13}
}

@article{Hinds.1998,
  title = {User Control and Its Many Facets: {{A}} Study of Perceived Control in Human-Computer Interaction},
  author = {Hinds, Pamela J.},
  year = 1998
}

@article{Inga.2023,
  title = {Human-Machine Symbiosis: {{A}} Multivariate Perspective for Physically Coupled Human-Machine Systems},
  author = {Inga, Jairo and Ruess, Miriam and Robens, Jan Heinrich and Nelius, Thomas and Rothfu{\ss}, Simon and Kille, Sean and Dahlinger, Philipp and Lindenmann, Andreas and Thomaschke, Roland and Neumann, Gerhard and Matthiesen, Sven and Hohmann, S{\"o}ren and Kiesel, Andrea},
  year = 2023,
  journal = {Int. Jour. of Human-Computer Studies},
  volume = {170},
  publisher = {Elsevier BV},
  doi = {10.1016/j.ijhcs.2022.102926}
}

@article{Javdani.2015,
  title = {Shared {{Autonomy}} via {{Hindsight Optimization}}},
  author = {Javdani, Shervin and Srinivasa, Siddhartha S. and Bagnell, J. Andrew},
  year = 2015,
  journal = {Robotics science and systems : online proceedings},
  volume = {2015},
  doi = {10.15607/RSS.2015.XI.032},
  urldate = {2026-02-25},
  pmcid = {PMC6329599},
  pmid = {30637295}
}

@article{Jenkins.2021,
  title = {An Investigation of "{{We}}" Agency in Co-Operative Joint Actions},
  author = {Jenkins, Michael and Esemezie, Olisaemeka and Lee, Vivian and Mensingh, Merani and Nagales, Kien and Obhi, Sukhvinder S.},
  year = 2021,
  journal = {Psychological Research},
  volume = {85},
  number = {8},
  doi = {10.1007/s00426-020-01462-6},
  pmid = {33398449}
}

@article{Johansson.2005,
  title = {Failure to {{Detect Mismatches Between Intention}} and {{Outcome}} in a {{Simple Decision Task}}},
  author = {Johansson, Petter and Hall, Lars and Sikstr{\"o}m, Sverker and Olsson, Andreas},
  year = 2005,
  journal = {Science},
  volume = {310},
  number = {5745},
  publisher = {American Association for the Advancement of Science},
  doi = {10.1126/science.1111709},
  urldate = {2026-02-12}
}

@article{Kaber.2004,
  title = {The Effects of Level of Automation and Adaptive Automation on Human Performance, Situation Awareness and Workload in a Dynamic Control Task},
  author = {Kaber, David B. and Endsley, Mica R.},
  year = 2004,
  journal = {Theoretical Issues in Ergonomics Science},
  volume = {5},
  number = {2},
  doi = {10.1080/1463922021000054335},
  urldate = {2025-02-27},
  langid = {english}
}

@inproceedings{Kille.2024,
  title = {Human-{{Variability-Respecting Optimal Control}} for {{Physical Human-Machine Interaction}}},
  booktitle = {2024 33rd {{IEEE International Conference}} on {{Robot}} and {{Human Interactive Communication}} ({{ROMAN}})},
  author = {Kille, Sean and Leibold, Paul and Karg, Philipp and Varga, Balint and Hohmann, S{\"o}ren},
  year = 2024,
  address = {Los Angeles},
  doi = {10.1109/RO-MAN60168.2024.10731297},
  urldate = {2025-12-17}
}

@misc{Kochhar.2023,
  title = {Which {{U}}.{{S}}. {{Workers Are More Exposed}} to {{AI}} on {{Their Jobs}}?},
  author = {Kochhar, Rakesh},
  year = 2023,
  journal = {Pew Research Center},
  urldate = {2025-03-21},
  langid = {american}
}

@incollection{Laugwitz.2008,
  title = {Construction and Evaluation of a User Experience Questionnaire},
  booktitle = {{{HCI}} and {{Usability}} for {{Education}} and {{Work}}},
  author = {Laugwitz, Bettina and Held, Theo and Schrepp, Martin},
  editor = {Holzinger, Andreas},
  year = 2008,
  volume = {5298},
  publisher = {Springer Berlin Heidelberg},
  address = {Berlin, Heidelberg},
  doi = {10.1007/978-3-540-89350-9},
  urldate = {2025-01-22},
  isbn = {978-3-540-89349-3 978-3-540-89350-9},
  langid = {english}
}

@article{Lorenzini.2023,
  title = {Ergonomic Human-Robot Collaboration in Industry: {{A}} Review},
  shorttitle = {Ergonomic Human-Robot Collaboration in Industry},
  author = {Lorenzini, Marta and Lagomarsino, Marta and Fortini, Luca and Gholami, Soheil and Ajoudani, Arash},
  year = 2023,
  journal = {Frontiers in Robotics and AI},
  volume = {9},
  doi = {10.3389/frobt.2022.813907},
  urldate = {2025-03-21},
  langid = {english}
}

@article{Macenski.2022,
  title = {Robot {{Operating System}} 2: {{Design}}, Architecture, and Uses in the Wild},
  shorttitle = {Robot {{Operating System}} 2},
  author = {Macenski, Steven and Foote, Tully and Gerkey, Brian and Lalancette, Chris and Woodall, William},
  year = 2022,
  journal = {Science Robotics},
  volume = {7},
  number = {66},
  publisher = {American Association for the Advancement of Science},
  doi = {10.1126/scirobotics.abm6074},
  urldate = {2026-01-07}
}

@article{Maeda.2017,
  title = {Probabilistic Movement Primitives for Coordination of Multiple Human--Robot Collaborative Tasks},
  author = {Maeda, Guilherme J. and Neumann, Gerhard and Ewerton, Marco and Lioutikov, Rudolf and Kroemer, Oliver and Peters, Jan},
  year = 2017,
  journal = {Autonomous Robots},
  volume = {41},
  number = {3},
  doi = {10.1007/s10514-016-9556-2},
  urldate = {2025-08-20},
  langid = {english}
}

@article{Medina.2015,
  title = {Synthesizing Anticipatory Haptic Assistance Considering Human Behavior Uncertainty},
  author = {Medina, Jose Ramon and Lorenz, Tamara and Hirche, Sandra},
  year = 2015,
  journal = {IEEE Transactions on Robotics},
  volume = {31},
  number = {1},
  doi = {10.1109/TRO.2014.2387571}
}

@article{Ozen.2021,
  title = {Promoting {{Motor Variability During Robotic Assistance Enhances Motor Learning}} of {{Dynamic Tasks}}},
  author = {{\"O}zen, {\"O}zhan and Buetler, Karin A. and {Marchal-Crespo}, Laura},
  year = 2021,
  journal = {Frontiers in Neuroscience},
  volume = {14},
  publisher = {Frontiers},
  doi = {10.3389/fnins.2020.600059},
  urldate = {2025-08-20},
  langid = {english}
}

@article{Padalkar.2025,
  title = {Towards Safe and Efficient Learning in the Wild: {{Guiding RL}} with Constrained Uncertainty-Aware Movement Primitives},
  shorttitle = {Towards Safe and Efficient Learning in the Wild},
  author = {Padalkar, Abhishek and Stulp, Freek and Neumann, Gerhard and Silv{\'e}rio, Jo{\~a}o},
  year = 2025,
  journal = {IEEE Robotics and Automation Letters},
  publisher = {IEEE},
  urldate = {2026-02-25}
}

@article{Parasuraman.2000,
  title = {A Model for Types and Levels of Human Interaction with Automation},
  author = {Parasuraman, R. and Sheridan, T.B. and Wickens, C.D.},
  year = 2000,
  journal = {IEEE Transactions on Systems, Man, and Cybernetics - Part A: Systems and Humans},
  volume = {30},
  number = {3},
  doi = {10.1109/3468.844354},
  urldate = {2025-02-27}
}

@article{Rheinberg.2003,
  title = {Die {{Erfassung}} Des {{Flow-Erlebens}}},
  author = {Rheinberg, Falko and Vollmeyer, Regina and Engeser, Stefan},
  year = 2003
}

@article{Rubino.2024,
  title = {Switched Kinematic and Indirect Force Control of a Motorized Ankle Orthosis to Perturb the Ankle Joint during Walking},
  author = {Rubino, Nicholas and Manchola, Miguel and Duenas, Victor H.},
  year = 2024,
  journal = {IFAC-PapersOnLine},
  volume = {58},
  number = {30},
  doi = {10.1016/j.ifacol.2025.01.159},
  urldate = {2025-03-20},
  langid = {english}
}

@article{Schmager.2025,
  title = {Understanding {{Human-Centred AI}}: A Review of Its Defining Elements and a Research Agenda},
  shorttitle = {Understanding {{Human-Centred AI}}},
  author = {Schmager, Stefan and Pappas, Ilias O. and Vassilakopoulou, Polyxeni},
  year = 2025,
  journal = {Behaviour \& Information Technology},
  volume = {44},
  number = {15},
  publisher = {Taylor \& Francis},
  doi = {10.1080/0144929X.2024.2448719},
  urldate = {2026-02-12}
}

@article{Schmidtler.2017,
  title = {A Questionnaire for the Evaluation of Physical Assistive Devices ({{QUEAD}}): {{Testing}} Usability and Acceptance in Physical Human-Robot Interaction},
  shorttitle = {A Questionnaire for the Evaluation of Physical Assistive Devices ({{QUEAD}})},
  author = {Schmidtler, Jonas and Bengler, Klaus and Dimeas, Fotios and {Campeau-Lecours}, Alexandre},
  year = 2017,
  journal = {2017 IEEE International Conference on Systems, Man, and Cybernetics (SMC)},
  publisher = {IEEE},
  address = {Banff, AB},
  doi = {10.1109/SMC.2017.8122720},
  urldate = {2024-07-30},
  isbn = {9781538616451}
}

@article{Scholz.1999,
  title = {The Uncontrolled Manifold Concept: Identifying Control Variables for a Functional Task},
  shorttitle = {The Uncontrolled Manifold Concept},
  author = {Scholz, J. P. and Sch{\"o}ner, Gregor},
  year = 1999,
  journal = {Experimental Brain Research},
  volume = {126},
  number = {3},
  doi = {10.1007/s002210050738},
  urldate = {2026-02-25},
  copyright = {http://www.springer.com/tdm},
  langid = {english}
}

@article{Synofzik.2013,
  title = {The Experience of Agency: An Interplay between Prediction and Postdiction},
  shorttitle = {The Experience of Agency},
  author = {Synofzik, Matthis and Vosgerau, Gottfried and Voss, Martin},
  year = 2013,
  journal = {Frontiers in Psychology},
  volume = {4},
  publisher = {Frontiers},
  doi = {10.3389/fpsyg.2013.00127},
  urldate = {2026-02-13},
  langid = {english}
}

@article{Tapal.2017,
  title = {The Sense of Agency Scale: A Measure of Consciously Perceived Control over One's Mind, Body, and the Immediate Environment},
  author = {Tapal, Adam and Oren, Ela and Dar, Reuven and Eitam, Baruch},
  year = 2017,
  journal = {Frontiers in Psychology},
  volume = {8},
  doi = {10.3389/fpsyg.2017.01552},
  pmid = {28955273}
}

@article{Todorov.2002,
  title = {Optimal Feedback Control as a Theory of Motor Coordination},
  author = {Todorov, Emanuel and Jordan, Michael I.},
  year = 2002,
  journal = {Nat. Neurosci.},
  volume = {5},
  number = {11},
  publisher = {Nature Publishing Group},
  doi = {10.1038/nn963},
  pmid = {12404008}
}

@article{Todorov.2004,
  title = {Optimality Principles in Sensorimotor Control},
  author = {Todorov, Emanuel},
  year = 2004,
  journal = {Nat. Neurosci.},
  volume = {7},
  number = {9},
  doi = {10.1038/nn1309},
  pmid = {15332089}
}

@article{Varga.2024,
  title = {Toward Adaptive Cooperation: Model-Based Shared Control Using Lq-Differential Games},
  author = {Varga, Balint},
  year = 2024,
  journal = {Acta Polytechnica Hungarica 2024},
  volume = {21},
  number = {10},
  doi = {10.12700/APH.21.10.2024.10.27}
}

@article{Wen.2015,
  title = {The Sense of Agency during Continuous Action: Performance Is More Important than Action-Feedback Association},
  shorttitle = {The Sense of Agency during Continuous Action},
  author = {Wen, Wen and Yamashita, Atsushi and Asama, Hajime},
  editor = {Sokolov, Alexander N.},
  year = 2015,
  journal = {PLOS ONE},
  volume = {10},
  number = {4},
  doi = {10.1371/journal.pone.0125226},
  urldate = {2025-02-17},
  langid = {english}
}

@article{Wen.2022,
  title = {The Sense of Agency in Perception, Behaviour and Human--Machine Interactions},
  author = {Wen, Wen and Imamizu, Hiroshi},
  year = 2022,
  journal = {Nature Reviews Psychology},
  volume = {1},
  number = {4},
  doi = {10.1038/s44159-022-00030-6}
}

@article{Zanatto.2021,
  title = {Human-Machine Sense of Agency},
  author = {Zanatto, Debora and Chattington, Mark and Noyes, Jan},
  year = 2021,
  journal = {International Journal of Human-Computer Studies},
  volume = {156},
  doi = {10.1016/j.ijhcs.2021.102716},
  urldate = {2026-02-16},
  langid = {english}
}

% Biography
%\bio{}
% Here goes the biography details.
%\endbio

%\bio{pic1}
% Here goes the biography details.
%\endbio

\end{document}